\documentclass[12pt]{JHEP3}

\usepackage{epsfig}
\usepackage{graphicx,amsmath,amssymb}
\usepackage{epsfig,multicol}
\allowdisplaybreaks

\usepackage{bbm,bm,amsmath,amssymb}
\def\blfootnote{\xdef\@thefnmark{}\@footnotetext}

\long\def\symbolfootnote[#1]#2{\begingroup%
\def\thefootnote{\fnsymbol{footnote}}\footnote[#1]{#2}\endgroup}

\newcommand\cY{\mathcal Y}
\DeclareMathOperator\lcm{lcm}
\usepackage{amsmath,amssymb}
\usepackage[latin1]{inputenc}

\newcommand{\al}{\alpha}

\newcommand{\ep}{\epsilon}

\newcommand{\ga}{\gamma}

\newcommand{\la}{\lambda}

\newcommand{\si}{\sigma}

\newcommand{\De}{\Delta}

\newcommand{\La}{\Lambda}

\newcommand{\ba}{\mathbf{a}}

\newcommand{\tP}{\tilde{P}}
\newcommand{\tS}{\tilde{S}}

\def\NN{\mathbb{N}}
\def\RR{\mathbb{R}}

\def\ZZ{\mathbb{Z}}

\renewcommand\SS{\mathbb{S}}
\newcommand{\cD}{{\mathcal D}}

\newcommand{\cO}{{\mathcal O}}

\newcommand{\pd}{\partial}
\newcommand\minus\backslash

\newcommand\lan\langle
\newcommand\ran\rangle

\newcommand{\I}{{\mathrm i}}
\newcommand{\e}{{\mathrm e}}

\renewcommand\leq\leqslant
\renewcommand\geq\geqslant
\newlength{\intwidth}

\renewcommand{\theenumi}{(\roman{enumi})}

\def\AdS{\mathrm{AdS}}
\newcommand{\bga}{\bar\gamma}

\newcommand{\be}{\begin{eqnarray}}
\newcommand{\ee}{\end{eqnarray}}
\newcommand{\ben}{\begin{eqnarray*}}
\newcommand{\een}{\end{eqnarray*}}

\newcommand{\bcent}{\begin{center}}
\newcommand{\ecent}{\end{center}}
\newcommand{\benum}{\begin{enumerate}}
\newcommand{\eenum}{\end{enumerate}}
\newcommand{\bdesc}{\begin{description}}
\newcommand{\edesc}{\end{description}}
\newcommand{\bitem}{\begin{itemize}}
\newcommand{\eitem}{\end{itemize}}
\newcommand{\bquote}{\begin{quote}}
\newcommand{\equote}{\end{quote}}
\newcommand{\bhalfp}{\begin{minipage}{0.45\textwidth}}
\newcommand{\ehalfp}{\end{minipage}}
\newcommand{\bhead}{\begin{center}\bf \Large}
\newcommand{\ehead}{\end{center}\bigskip}

 \newcommand{\bg}{\begin{equation}}
 \newcommand{\nd}{\end{equation}}

\def\be{\begin{equation}}
\def\ee{\end{equation}}
\def\ba{\begin{eqnarray}}
\def\ea{\end{eqnarray}}

\newcommand{\roughly}[1]{\mathrel{\raise.3ex\hbox{$#1$\kern-0.85em
\lower1ex\hbox{$\sim$}}}}

\def\2pi{\left(2\pi\right)}

\def\minus{$-$}

\def\beq{\begin{equation}}
\def\eeq{\end{equation}}
\def\bg{\begin{eqnarray}}
\def\nd{\end{eqnarray}}
\def\bea{\begin{eqnarray}}
\def\eea{\end{eqnarray}}

\def\D3{\overline{\mbox{D3}}}

\def\cD{{\cal D}}

\newcommand{\tw}{{\widetilde w}}

\title{On the Scalar Spectrum of the $Y^{p,q}$ Manifolds}

\author{Fang Chen${}^1$, Keshav Dasgupta${}^1$, Alberto Enciso${}^2$, Niky Kamran${}^3$, Jihye Seo${}^1$\\
\vskip.03in
${}^1$ Ernest Rutherford Physics Building, McGill University,\\
3600 University Street, Montr{\'e}al QC, Canada H3A 2T8\\
${}^2$ Instituto de Ciencias Matem\'aticas\\
Consejo Superior de Investigaciones Cient\'\i ficas, 28049 Madrid, Spain\\
${}^3$ Department of Mathematics and Statistics, McGill University,\\
805 Sherbrooke Street West, Montr{\'e}al QC, Canada H3A 2K6\\
{\tt fangchen, keshav, jihyeseo@hep.physics.mcgill.ca, nkamran@math.mcgill.ca, aenciso@icmat.es}}
\date{October 2011}

\abstract{The spectra of supergravity modes in anti de Sitter (AdS) space on a five-sphere endowed with the round metric (which is the simplest
5d Sasaki-Einstein space) has been studied in detail in the past. However for the more general class of cohomogeneity one Sasaki-Einstein
metrics on $S^2\times S^3$, given by the $Y^{p, q}$ class, a complete study of the spectra has not been attempted. Earlier studies on scalar
spectrum were restricted to only the first few eigenstates.
In this paper we take a step in this direction
by analysing the full scalar spectrum on these spaces.
However it turns out that finding the exact solution of the corresponding eigenvalue
problem in closed form is not feasible since the computation of the eigenvalues of the Laplacian boils down to the analysis of a one-dimensional
operator of Heun type, whose spectrum cannot be computed in closed form. However, despite this analytical obstacle,
we manage to get both lower and upper bounds on the
eigenvalues of the scalar spectrum by comparing the eigenvalue problem with a simpler, solvable system. We also briefly touch upon various other
new avenues such as non-commutative and dipole deformations as well as possible non-conformal extensions of these models.}

\maketitle

\begin{document}

\section{Introduction}

The gravity dual of ${\cal N} = 1$ CFT has been studied earlier from many different perspectives starting with \cite{klebwit} where the associated
CFT, endowed with a simple product gauge group and a simple quartic superpotential, appeared from $N$ D3-branes placed at the tip of a conifold
geometry.
One way to change the gauge group and the
superpotential structure is to change the underlying
conifold geometry itself by either an orbifolding or an orientifolding action. A subsequent T-duality,
 mapping these actions to either the {\it interval} \cite{uranga, Dasgupta:1998su} or the {\it brane-box} models \cite{BB1, BB2}, then
gives us simple ways to analyse the underlying ${\cal N} = 1$ CFTs.

An alternative way to change the gauge group and the superpotential structure is to change the Calabi-Yau condition of the conifold itself, namely,
change the K\"ahler class and the complex structures so as to put different Ricci flat metrics on the conifold.
Since there are infinite ways of
doing it, there would exist infinite variations of the conifold that are all Calabi-Yau manifolds. All of these would lead to gravity duals
of the form $AdS_5 \times Y^{p, q}$ where $Y^{p, q}$ are the so-called  Sasaki-Einstein manifolds. These ideas, including the 
underlying gauge/gravity duality, were developed few years ago in \cite{0402153, Sp04, MS1, MS2}. 

In this paper we study spectrum of Sasaki-Einstein manifold $Y^{p,q}$, using spectral-theoretic methods, continuing the work of \cite{EK10}.
More precisely,
we study the Laplacian operator of a $Y^{p,q}$ manifold, associated to its scalar spectrum, using the framework laid out in \cite{EK10}.
The authors of
\cite{EK10} analyzed the Cauchy problem, and presented a Fourier-type decomposition for the eigenfunction.
In order to use spectral-theoretic methods, they used the Friedrichs extension of the Laplacian operator to
rule out logarithmic singularities. This way a self-adjoint extension of unbounded symmetric operator could be determined. Our starting point, in this
paper,
is to use this operator to study its eigenmodes.

The
{\it lowest} eigenmodes of the Laplacian were first
studied in \cite{Kihara:2005nt} for $Y^{p,q}$, wherein they also tried to construct an AdS/CFT dictionary. This work was
followed by \cite{Oota:2005mr} where they studied the lowest eigenmodes for more generic manifolds like the
$L^{a,b,c}$ examples. An important progress in \cite{Kihara:2005nt} was the realization that
the Laplacian operator could be expressed in terms of a Heun type operator, whose lowest modes are easily computable.
However, for higher modes not much progress has been made in the literature.
Even numerical studies do not look simple.
In \cite{Bachas:2011xa}, the spectrum is studied numerically for $S^5$ case,
which is the simplest Sasaki-Einstein manifold in 5d, but an equivalent work for the $Y^{p, q}$ case is still lacking.

In this paper, we will use mathematical tools developed in analysis and spectral theory, to address the question of finding all the eigenmodes.
However, as it will turn out, finding the exact solution of the eigenvalue problem in closed form does not seem feasible since the
computation of the eigenvalues of the Laplacian boils down to the analysis of a one-dimensional differential operator (of Heun type),
which has {\em four}\/
regular singular points. What we will do, therefore, is to find {\it bounds} for the eigenvalues of the operator,
which will allow us to approximate the conformal dimensions of the dual CFTs. In our approach,
we will get results in two different regimes, which match in their overlap. The first is for highly excited modes
(by focussing on the leading terms in mode number $k$), and the second is for small $a$,
where $0<a<1$ parametrizes  $Y^{p,q}$ geometry by implicitly parametrizing ($p, q$). The parameter $a$ is
determined by $p,q$, and $a \ll 1$ is equivalent to $q \ll p$.

Our work uses some techniques in analysis and spectral theory that may not be too familiar to some readers in the physics community.
Additionally, a more physical motivation to
study Sasaki-Einstein manifolds is never spelled out in the literature, although detailed mathematical reviews exist.
Therefore
in the following sub-sections we will introduce two concepts to the reader.
First will be the motivation to study Sasaki-Einstein manifolds in general; and the
second will be the minimal mathematical background necessary to sketch the mathematical techniques used in this paper.

\subsection{Motivation to study Sasaki-Einstein manifolds}

Here we will introduce and motivate the study of Sasaki-Einstein manifolds, mostly summarizing the extensive reviews in \cite{SparksReview, 1004.2461} in a  language slightly more appropriate for the physicists.
A Sasaki-Einstein manifold is both Sasakian and Einstein, and it is an odd-dimensional cousin of K\"ahler-Einstein manifold, and sandwiched between
two K\"ahler-Einstein manifolds of one dimension lower and higher respectively \cite{SparksReview}. Einstein condition of
Sasaki/K\"ahler manifold is inherited between lower and higher dimensions (see \cite{Sfetsos:2005kd} for examples for $d = 4, 5, 6$).

Two mathematical facts about these manifolds are readily available:
Cone over Sasaki-Einstein is K\"ahler-Einstein manifold with one dimension higher; and
K\"ahler geometry is a symplectic geometry, while Sasakian is a contact geometry.
Extra physical motivation comes from the fact that in Hamiltonian mechanics phase space with $n$ momentum and $n$ position forms a $2n$ dimensional symplectic manifold. By adding one more direction, i.e the time evolution, we get $2n+1$ dimensional contact manifold. Contact geometry (therefore also Sasakian geometry) is just as important as symplectic geometry for physicists, as one can see in \cite{Dahl} for example. Extensive details on contact geometry are given in the handbook \cite{2003math......7242G} for the enthusiastic readers to dwell upon. 

Another property of these manifolds is associated to their Reeb vectors.
If, for example, the
Reeb vector fields have compact orbits forming circles and if the $U(1)$ actions are free (not free, resp.), then the Sasakian manifolds are
regular (quasi-regular, resp.). If, on the other hand, the orbits are non-compact, the Sasakian manifolds are irregular.

In contrast to the fact that there are abundant four and six-dimensional K\"ahler-Einstein manifolds, until recently there were only
two known Sasaki-Einstein manifolds in 5d, namely the $S^5$ and $T^{1,1}$.
Using the M-theory solution of \cite{0402153}, the authors of
\cite{Sp04} found examples of new
5d Sasaki-Einstein metrics $Y^{p,q}$ on $S^2 \times S^3$. These examples
contain both quasi-regular and regular cases, and the corresponding CFT duals have rational and irrational central charges 
respectively \cite{MS1, cotrone1, MS2}.
These
Sasaki-Einstein metrics are also critical points of volume functional \cite{SparksReview,0603021}. Note that
a bigger group of 5d Sasaki-Einstein manifolds namely the $L^{abc}$ manifolds contain the $Y^{p,q}$ manifolds as a subset. The metric for this bigger
class of manifolds
were constructed from Kerr Black hole solutions by taking some scaling limits (see \cite{Cvetic} for more details).

Finally, another motivation to study these manifolds will be to
build brane constructions in flat space which are T-duals to the $Y^{p,q}$ geometries (much like the one for the $T^{1,1}$ case in
\cite{uranga, Dasgupta:1998su}). This will not only help us to analyze the corresponding gauge theories but will also provide new brane constructions
in string theory.


\subsection{A brief sketch of the mathematical techniques for the physicists}

After having discussed the physical motivations to study the $Y^{p, q}$ manifolds, let us summarise the key mathematical concepts that we will be using
throughout the paper, i.e the concept of unbounded operators and Friedrichs extensions. For completeness we will also give a brief discussion of
the Sturm-Liouville theory.


\subsubsection{Quantum mechanical observables and unbounded operators}

As in many parts of quantum mechanics, unbounded self-adjoint
operators will play an important role in this paper. For the benefit of the
reader, we will briefly review some basic ideas that we will touch
upon later. We recall that a linear operator $L$ between two normed
vector spaces $X$ and $Y$ is {\em bounded}, or continuous, if  the ratio of the norm
of $Lv$ to that of $v$ remains bounded. We will be mainly interested
in the case when $X=Y$ is a Hilbert space, usually some $L^2$ space. It is well known that linear
operators between finite-dimensional vector spaces are always bounded.

A bounded linear  operator is self-adjoint if and only if it is
symmetric (i.e., Hermitian). For unbounded operator, this is not the
case: there are examples of unbounded symmetric operators which are
not self-adjoint, due to subtleties regarding the domain of the
operator. Since self-adjointness (and {\em not} mere symmetry) is key
for the validity of the spectral theorem, for the purposes of quantum
mechanics it is often crucial to ensure that a given unbounded
operator is self-adjoint with a given domain of
definition. We remark as well that the domain of an unbounded operator
can never be the whole Hilbert space, but only dense in it. (Incidentally, let us remark that observables in
quantum mechanics, including the free Hamiltonian in $\RR^d$, the
Coulomb Hamiltonian, and the position, momentum and angular momentum
operators, are unbounded, self-adjoint operators, and this was the
motivation for von Neumann and M.\ Stone's original work in this area.)

The need to have bona fide self-adjoint operators leads to the theory
of self-adjoint extensions. Given a symmetric operator densely defined
in a Hilbert space, it does not necessarily admit a self-adjoint
extension, and even when it does, this extension does not need to be
unique, and deciding which extension is physically relevant is
nontrivial. Fortunately for us, in this work all the self-adjoint
extensions we shall need are of Friedrichs type, which is the
preferred, time-honored way to define self-adjoint extensions of
lower-bounded operators. For our purposes, it is enough to know that
the Friedrichs extension is a standard procedure to derive a
self-adjoint operator, densely defined in an $L^2$ space, from an
operator $L$ whose action in the set of test functions is lower
bounded (that is, $\langle Lv, v\rangle \geq -C \|v\|^2$ for all $v\in C^\infty_0$). The
idea of this method is that $L$ can be used to define a stronger
norm (in quantum mechanics, typically of Sobolev type) which allows to
complete the minimal domain of $L$ to get a larger domain in which the operator
is self-adjoint. This extension is widely used in physics; for example,
it is the usual way to define operators with Dirichlet boundary conditions.

\subsubsection{Sturm-Liouville theory}

A one-dimensional Sturm-Liouville operator of second-order is of the form
\bg\label{SL}
  Lf(t)=\frac1{W(t)}\Bigg[-\frac d{d t}\Big(P(t)\frac{d f(t)}{d
  t}\Big)+Q(t)f(t)\Bigg]\,,
\nd
with $W,P$ nonnegative functions in an interval of the real line $(a,b)$. The importance of Sturm--Liouville
operators is that it is a class of symmetric operators for which we
have a lot of information about their self-adjointness and spectra.  In particular, and depending
on the properties of the functions $W,P,Q$ that define the operator
and its domain of definition, sometimes we have formulas for the
essential spectrum of the operator or for the asymptotic value of its
eigenvalues. We will make use of some of them in forthcoming sections
but, as technical conditions are sometimes hard to express without a
concrete example in view, we will refrain from stating them at this
point. We recall that  special functions, such as
Bessel functions or Laguerre polynomials, are often
defined as solutions to the eigenvalue problem of a Sturm-Liouville
operator.

\subsection{Organization of the paper}

The paper is organized in the following way.
In section \ref{YpqReview} we review the basics of $Y^{p,q}$ geometry.
Section \ref{YpqSpec} studies the scalar spectrum of $Y^{p,q}$ geometry by analysing the solution of the Laplacian operator.
In subsection \ref{harExp},
we present how the spectrum of Type IIB on $AdS_5 \times Y^{p,q}$ is related to the scalar spectrum of $Y^{p,q}$ geometry.
In subsection \ref{ScalarY}
scalar modes in $Y^{p,q}$ are studied, by separating the variables in the wave function for the scalar modes.
Behaviour of the eigenvalues for highly excited modes is studied in subsection \ref{largek} using Sturm-Liouville theory. In subsection \ref{smalla}, we compare Laplacian operator with simpler solvable operators in order to give upper and lower bounds for the all eigenvalues, which works best for
$a \ll  1$ or equivalently $q \ll p$.
Section \ref{example} discusses examples of various other modes and analyze cases that may take us beyond the scalar spectra of IIB. Subsection
\ref{IIAbrane} studies possible type IIA brane realisation, and subsection \ref{dipole} discusses non-commutative and dipole deformations. 
One may note that in this section (and also the next) we will not address the spectra of the theory. To analyse the spectra we would not only need to go beyond the 
scalar fields, but would also require exact eigenvalues of the KK modes for all spin-states of the theory $-$ a calculation that will be relegated for future works. 
In
section \ref{geomtr} we go beyond the conformal cases to
study new non-conformal duals that may arise from
possible geometric transitions. Earlier results in this direction were more along the lines of cascading theories of Klebanov-Strassler type. To
study geometric transitions for our case, we need both the resolved and the deformed cone over the $Y^{p, q}$ manifolds.
In subsection \ref{resolY}, we review the metric for the cone $Y^{p,q}$ after resolution and then discuss the possibility of generating deformed
cones over $Y^{p, q}$ base. We briefly argue why these deformations may not give rise to K\"ahler or complex manifolds. In
subsection \ref{D5resolY} we
discuss the first step of geometric transitions, namely, constructions of the backgrounds with wrapped
D5 branes on the resolved cones over the $Y^{p,q}$ manifolds. In subsection \ref{GT} we discuss the actual process of geometric transitions briefly
and point our possible issues that may make the underlying calculations highly non-trivial. Finally in
\ref{conclusion} we conclude by pointing out various future directions.
In appendix \ref{eigenAppen}, eigenvalues of the differential operator $S$ (Laplacian) are discussed, mostly borrowing some results from \cite{EK10}.

\section{$Y^{p,q}$ geometry \label{YpqReview}}
The $Y^{p,q}$ metrics are Sasaki-Einstein and therefore a cone over
them is Calabi-Yau. We start with the local metric
\begin{eqnarray}\label{stdm}
ds^2&=&\frac{1-cy}{6}(d\theta^2+\sin^2\theta
d\phi^2)+\frac{1-cy}{2f(y)}dy^2+\frac{f(y)}{9(a-y^2)}(d\psi-\cos\theta
d\phi)^2\nonumber\\
&&+\frac{2(a-y^2)}{1-cy}\Big[d\alpha+\frac{ac-2y+y^2c}{6(a-y^2)}(d\psi-\cos\theta
d\phi)\Big]^2
\end{eqnarray}
where $f(y)=2cy^3-3y^2+a$. As in \cite{Sp04} one can
show that $Ric=4g$ for all values of $a$ and $c$ therefore satisfying Einstein condition. For $c=0$ and
$a=3$ the metric is exactly the local form of the standard metric on
$T^{1,1}$. For $c\neq 0$ one can always rescale $y$ ($y \rightarrow y/c$, and also $a \rightarrow a/c^2$, $f\rightarrow f/c^2$, etc) to set $c=1$
which we will take in the following.

It is obvious that the first two terms give the metric of an $S^2$
for a fixed $y$, if the periodicity of $\theta$ and $\phi$ are $\pi$
and $2\pi$ respectively. To study the ($y$, $\psi$) space one first
requires
\begin{eqnarray}
&&1-y>0,\quad a-y^2>0\nonumber\\
&&f=a-3y^2+2y^3\geq0.
\end{eqnarray}
In order for $y$ to have solutions $a$ must satisfy $0<a<1$. The negative solution of $f=0$ and the
smallest positive solution are denoted by $y_-$ and  $y_+$ respectively. Then $y$ needs to take values between $y_-<y<y_+$
, (so that all the terms in the metric come with positive sign). When $a=1$ the metric \eqref{stdm} is the
local round metric of $S^5$. If $\psi$ has the period of $2\pi$ then
($y$, $\psi$) is topologically a 2-sphere\footnote{The range of $y$ is taken to be $[y_-,y_+]$. This ensures that $w$
(defined in \eqref{defop})
is strictly positive in this interval and $r\geq 0$, vanishing only
at the endpoints $y_\pm$. If we identify $\psi$ periodically, the part
of $g_B$ ($g_B$ is only defined in \cite{EK10} but not in this paper) given by
$$\frac{1-cy}{2f(y)} ~dy^2 + \frac{f(y)}{9(a-y^2)}~d\psi^2$$
describes a circle fibered over the interval $(y_-,y_+)$, the size
of the circle shrinking to zero at the endpoints. Remarkably, the
$(y,\psi)$ fibers are free of conical singularities if the period of
$\psi$ is $2\pi$, in which case the circles collapse smoothly and
the $(y, \psi)$ fibers are diffeomorphic to a $2$-sphere.}.

In order to have a compact manifold one takes the period of $\alpha$
to be $2\pi l$. Then $l^{-1}A$, where $A$ is the last term in the
second line of \eqref{stdm}, becomes a connection on a $U(1)$
bundle over $S^2\times S^2$ which puts constraints on $A$. In general
such $U(1)$ bundles are completely specified topologically by the
gluing on the equator of the two $S^2$ cycles, $C_1$ and $C_2$.
These are measured by the corresponding Chern numbers in
$H^2(S^2, \ZZ)= \ZZ$ which will be labeled as $p$ and $q$. The Chern
numbers are given by the integrals of $l^{-1}A/2\pi$ over $C_1$ and
$C_2$, namely:
\bg
p ~= ~ \frac{1}{2\pi l}\int_{C_1} A=\frac{y_--y_+}{6y_-y_+}, \quad \quad
q ~= ~ \frac{1}{2\pi l}\int_{C_2} A=\frac{(y_--y_+)^2}{9y_-y_+}
\nd
From their ratio ${p\over q}={3\over 2(y_+-y_-)}$, it follows
\bg
a ~= ~ \frac{1}{2}-\frac{p^2-3q^2}{4p^3}\sqrt{4p^2-3q^2}, \quad \quad
l ~= ~\frac{q}{3q^2-2p^2+p\sqrt{4p^2-3q^2}}
\nd
Metric \eqref{stdm} can be written in a canonical way if one makes the
coordinate change
\begin{equation} \label{newcoor}
\alpha=-\beta/6-c\psi^{\prime}/6, \quad \quad \psi=\psi^{\prime}\end{equation}
 to \eqref{stdm}. This converts \eqref{stdm} to the
following metric:
\begin{eqnarray}\label{canm}
ds^2&=&\frac{1-cy}{6}(d\theta^2+\sin^2\theta
d\phi^2)+\frac{1-cy}{2f(y)}dy^2+\frac{f(y)}{18(1-cy)}(d\beta+c\cos\theta
d\phi)^2\nonumber\\
&&+\frac{1}{9}(d\psi-\cos\theta d\phi+y(d\beta+c\cos\theta d\phi))^2.
\end{eqnarray}
The Killing vector
\begin{eqnarray}
\frac{\partial}{\partial\psi^{\prime}}=\frac{\partial}{\partial\psi}-\frac{1}{6}\frac{\partial}{\partial\alpha}
\end{eqnarray}
is globally well defined. For a generic 
value of $a$ its orbit is
not closed, in which case the Sasaki-Einstein metric is irregular. It is quasi-regular, if and only if $4p^2-3q^2=m^2, m\in \mathbb{Z}$.

\section{The spectrum of the $Y^{p,q}$ manifolds \label{YpqSpec}}

After our brief discussion of the geometry of the $Y^{p, q}$ manifolds, let us come to the main analysis of paper: the study of scalar spectrum of these manifolds.  We will start by analysing the solution of the Laplacian operator arising from the Fourier decomposition of functions as discussed earlier in~\cite{EK10}. However, as it will turn out, finding the exact solution of the eigenvalue problem in closed form does not seem feasible since the computation of the eigenvalues of the Laplacian boils down to the analysis of a one-dimensional differential operator (which we call $S$) of Heun type, which has {\em four}\/ regular singular points. What we will do, therefore, is to find bounds for the eigenvalues of $S$, which will allow us to approximate the conformal dimensions of the theory. In subsection \ref{example}, we will study some examples of these modes and discuss cases that may take us beyond the scalar spectra.

\subsection{Harmonic expansion on $Y^{p,q}$ \label{harExp}}
We will follow the argument in \cite{Ceresole:1999ht} which gives
the spectrum of Type IIB on $AdS_5\times T^{1,1}$. The background
solution in Type IIB is
\begin{eqnarray}
ds^2=\frac{r^2}{R^2}(-dx_0^2+dx_i^2)+\frac{R^2}{r^2}dr^2+R^2ds^2_{Y^{p,q}}
\end{eqnarray}
with the self-dual 5-form flux $F_5=(1+\ast)dx_0\wedge dx_1\wedge
dx_2\wedge dx_3\wedge d\Big(\frac{r^4}{R^4}\Big)$.

When Kaluza$-$Klein reducing this solution to $AdS_5$, we first have
to compute the fluctuations of the $10$-dimensional fields. The
fluctuation of the gravitational fields are parametrized as
\begin{eqnarray}
\tilde{g}_{\mu\nu}=g_{\mu\nu}+h_{\mu\nu}-\frac{1}{3}g_{\mu\nu}h^{a}_{a}, \quad
\,\, \tilde{g}_{\mu a}=h_{\mu a}, \quad  \,\, \tilde{g}_{ab}=g_{ab}+h_{ab}
\end{eqnarray}
where $\mu$, $\nu$ denote the $AdS_5$ space time while $a$, $b$
denote the internal space, and $g$ denotes the background metric
while $h$ is the fluctuation.

Now we expand the fields $h_{\mu\nu}$, $h_{\mu a}$, $h_{a b}$ and
$h^a_a$ into a complete set of harmonic functions on $Y^{p,q}$. With
the de Donder and Lorentz-type gauge conditions $D^ah_{(ab)}=0$ and
$D^ah_{a\mu}=0$ we have the following expansions\footnote{($x, y$) denote coordinates of the $AdS_5$ and $Y^{p,q}$ spaces respectively and therefore
should not be confused with the $y$ coordinates that we will be using to write the metric etc of the $Y^{p,q}$ spaces \label{ft2}.}:
\begin{eqnarray}\label{decomp1}
h_{\mu\nu}(x,y)&=&\sum_{\{\lambda\}} H^{\{\lambda\}}_{\mu\nu}(x)Y^{\{\lambda\}}(y),\quad h^a_a(x,y)=\sum_{\{\lambda\}} \pi^{\{\lambda\}} (x)Y^{\{\lambda\}}(y)\nonumber\\
h_{a\mu}(x,y)&=&\sum_{\{\lambda\}} B^{\{\lambda\}}_{\mu}(x)Y^{\{\lambda\}}_{a}(y),\quad h_{(ab)}(x,y)=\sum_{\{\lambda\}} \phi^{\{\lambda\}}(x)Y^{\{\lambda\}}_{(ab)}(y)\nonumber\\
\end{eqnarray}
where $[\lambda]\equiv [\lambda_1,\cdots,\lambda_{[5/2]}]$ denotes the
SO(5) representation.
Similarly with the gauge condition $D^aA_{a\mu}=0$ and $D^aA_{ab}=0$
we can expand the type IIB complex zero and the two-forms, $B$ and $A_{mn}$ respectively, as
\begin{eqnarray}\label{decomp2}
A_{\mu\nu}(x,y)&=&\sum_{\{\lambda\}}a^{\{\lambda\}}_{\mu\nu}(x)Y^{\{\lambda\}}(y),\quad
A_{a\mu}(x,y)=\sum_{\{\lambda\}}a^{\{\lambda\}}_{\mu}(x)Y^{\{\lambda\}}_a(y)\nonumber\\
A_{ab}(x,y)&=&\sum_{\{\lambda\}}a^{\{\lambda\}}(x)Y_{[ab]}^{\{\lambda\}}(y),\quad
B(x,y)=\sum_{\{\lambda\}}B^{\{\lambda\}}(x)Y^{\{\lambda\}}(y)
\end{eqnarray}
For the four-form flux we can do the same thing by imposing the
conditions $D^aa_{abcd}=0$, $D^aa_{abc\mu}=0$, $D^aa_{ab\mu\nu}=0$
and $D^aa_{a\mu\nu\gamma}=0$,
\begin{eqnarray}\label{decomp3}
a_{abcd}&=&\sum_{\{\lambda\}}b^{\{\lambda\}}(x)Y^{\{\lambda\}}_{abcd}(y),\quad
a_{abc\mu}=\sum_{\{\lambda\}}b^{\{\lambda\}}_{\mu}(x)Y^{\{\lambda\}}_{abc}(y)\nonumber\\
a_{ab\mu\nu}&=&\sum_{\{\lambda\}}b^{\{\lambda\}}_{\mu\nu}(x)Y^{\{\lambda\}}_{ab}(y),\quad
a_{a\mu\nu\gamma}=\sum_{\{\lambda\}}b^{\{\lambda\}}_{\mu\nu\gamma}(x)Y^{\{\lambda\}}_{a}(y)\nonumber\\
a_{\mu\nu\gamma\rho}&=&\sum_{\{\lambda\}}b^{\{\lambda\}}_{\mu\nu\gamma\rho}(x)Y^{\{\lambda\}}(y)
\end{eqnarray}
Notice that $Y^{p,q}$ is topologically $S^2\times S^3$, the same as
$T^{1,1}$, so we can argue similarly as in \cite{Ceresole:1999ht} to
simplify the expansion,
\begin{eqnarray}\label{decomp4}
a_{abcd}=\sum_{\{\lambda\}}b^{\{\lambda\}}(x)\epsilon_{abcd}^eD_eY^{\{\lambda\}}(y)
\end{eqnarray}
The full linearlized equation of motion can be found in
\cite{Kim:1985ez}. In this paper we are only interested in scalar harmonics which means that we are only looking at the
following modes in $AdS_5$, coming from first line of \eqref{decomp1}, \eqref{decomp2}, and the last line of \eqref{decomp3}:
\begin{eqnarray}\label{decomp5}
&& h_{\mu\nu}(x,y) = \sum_{\{\lambda\}} H^{\{\lambda\}}_{\mu\nu}(x)Y^{\{\lambda\}}(y), ~~~~~ h^a_a(x,y)=\sum_{\{\lambda\}} \pi^{\{\lambda\}} (x)
Y^{\{\lambda\}}(y)\nonumber\\
&& A^{(i)}_{\mu\nu}(x,y) = \sum_{\{\lambda, i\}}a^{\{\lambda\}}_{\mu\nu}(x)Y^{\{\lambda\}}(y), ~~~~~
B^{(j)}(x,y)=\sum_{\{\lambda\}}B^{\{\lambda, j\}}(x)Y^{\{\lambda\}}(y)\nonumber\\
&& a_{\mu\nu\gamma\rho}=\sum_{\{\lambda\}}b^{\{\lambda\}}_{\mu\nu\gamma\rho}(x)Y^{\{\lambda\}}(y)
\end{eqnarray}
where $A^{(i)}_{\mu\nu}(x,y)$ would be the NS and RR two-forms respectively and $B^{(j)}(x,y)$, where $i,j=1,2$,
would be the axion and the dilaton respectively.
The
other two quantities $\pi$ and $b$ that appear respectively from the expansion of $h^a_a$ in \eqref{decomp1} and from the expansion of $a_{abcd}$ in
\eqref{decomp4}, are related to the metric and the four-form respectively.
Therefore taking all these into account, we
are left with the following equations:
\begin{eqnarray}
&&(\square_x+\boxtimes_y)H_{\mu\nu}^{\{\lambda\}}=0\nonumber\\
&&(\square_x+\boxtimes_y)B^{\{\lambda, j\}}=0\nonumber\\
&&(\textrm{Max}+\boxtimes_y)a_{\mu\nu}^{\{\lambda, i\}}+\frac{2i}{R}
\epsilon_{\mu\nu}^{\quad\sigma\tau\gamma}\partial_{\sigma}a_{\tau\gamma}^{\{\lambda, i\}}=0\nonumber\\
 &&\square_x \Big(\begin{array}{clcr}
 \pi^{\{\lambda\}}\\
 b^{\{\lambda\}}
 \end{array}\Big)+\Big(\begin{array}{clcr}
 &\boxtimes_y-32R^{-2} & \quad 80R^{-1}\boxtimes_y\\
 &-\frac{4}{5}R^{-1} & \boxtimes_y
 \end{array}\Big)
 \Big(\begin{array}{clcr}
 \pi^{\{\lambda\}}\\
 b^{\{\lambda\}}
 \end{array}\Big)=0\nonumber\\
 \end{eqnarray}
where Max denotes the Maxwell operator and $\square_x$, $\boxtimes_y$ are the kinetic operators in the
$AdS_5$ space time and $Y^{p,q}$
spaces respectively. In our case the latter is exactly given by the action of the covariant
Laplacian operator
on the corresponding $SO(5)$ representation\footnote{For more details on the Maxwell and the Laplacian operator see \cite{Kim:1985ez, Ceresole:1999ht}.},
which can be formally written as
\begin{eqnarray}
\boxtimes_y \equiv {\square_y Y^{\{\lambda\}}\over Y^{\{\lambda\}}}
\end{eqnarray}
Our next step then is to analyze the eigenvalues of the Laplacian operator in
order to find the mass spectrum for these fields.

\subsection{Scalar modes in $Y^{p, q}$ \label{ScalarY}}

As we discussed in detail in the above subsection,
our goal is to compute the eigenvalues $\la_n$ of the Laplacian in the
manifold $Y^{p,q}$. These eigenvalues enter the scalar wave equation
on $\AdS_5\times Y^{p,q}$ as masses, so that the conformal dimensions
of the associated fields at infinity (i.e for the CFT dual)
are given by Witten's formula \cite{witten}:
\[
\De_k= 2 + \sqrt{4+\la_k}\,.
\]
It is well known that the Laplacian on $Y^{p,q}$, which we denote by
$\square_y$, defines a nonnegative, self-adjoint operator whose domain is
the Sobolev space $H^2(Y^{p,q})$ of square-integrable functions with
square-integrable second derivatives. The Laplacian is given in
local coordinates as \cite{EK10}\footnote{Note that $y$ denotes different things on LHS and RHS of \eqref{lapladef}. See footnote \ref{ft2}.}:
\begin{eqnarray}\label{lapladef}
\square_y~ \equiv ~  && g^{ij}\nabla_i\nabla_j=\frac{1}{\rho(y)}\frac\pd{\pd y}\rho(y)\,w(y)\,r(y)\,\frac\pd{\pd y} +\frac{1}{w(y)}\frac{\pd^2}{\pd\al^2}+ \frac{9}{r(y)}\bigg(\frac\pd{\pd\psi}-h(y)\,\frac\pd{\pd\al}\bigg)^2\nonumber\\
   && +\frac{6}{1-y}\Bigg[\frac1{\sin\theta}\frac\pd{\pd\theta}\sin\theta\frac\pd{\pd\theta} +\frac{1}{\sin^2\theta}\bigg(\frac\pd{\pd\phi}+\cos\theta\frac\pd{\pd\psi}\bigg)^2\Bigg]
\end{eqnarray}
where the various coefficients appearing above can be identified from \eqref{stdm} after rescaling to set $c=1$, i.e.,
\begin{equation}\label{defop}
w(y) \equiv \frac{2(a-y^2)}{1-y}, ~~
r(y) \equiv \frac{2y^3 - 3y^2 + a}{a-y^2}, ~~ h(y) \equiv \frac{y^2 - 2y + a}{6(a-y^2)}, ~~\rho(y) \equiv \frac{1-y}{18}
\end{equation}
The scalar mode $\Phi(y, \theta, \phi, \psi, \alpha)$ in the internal space now takes the following wave-functional form that was derived in \cite{EK10}:
\bg\label{scalarmode}
\Phi = u(y) v(\theta) e^{i(n\phi + 2m\psi + l\sigma\alpha/\tau)}
\nd
which means that the Laplacian satisfies:
\bg\label{laplasat}
\square_y \Phi ~ = ~ \left[S_{nmlj} u(y)\right] v(\theta) e^{i(n\phi + 2m\psi + l\sigma\alpha/\tau)}
\nd
where we saw in~\cite{EK10} that the analysis of the eigenvalues of the
Laplacian on $Y^{p,q}$ is reduced to that of (the Friedrichs extension
of) the one-dimensional
operators
\bg\label{jbaaz}
S_{nmlj}&~ \equiv ~& -\frac1{\rho(y)}\frac\pd{\pd y}\rho(y)\,w(y)\,r(y)\,\frac\pd{\pd y} +\frac1{w(y)}\bigg(\frac{\si l}\tau\bigg)^2+ \frac9{r(y)}\bigg(2m-h(y)\,\frac{\si l}\tau\bigg)^2
+\frac{6\La_{nmj}}{1-y}\,,\nonumber\\
& ~ =~ &-\frac2{1-y}\frac\pd{\pd y}(a-3y^2+2y^3) \frac\pd{\pd y}
+\frac{\ga^2(1-y)}{4(a-y^2)} +\frac{6\La_{nmj}}{1-y}\\
&& \qquad\qquad\qquad \qquad\qquad\qquad \qquad +\frac{9(a-y^2)}{a-3y^2+2y^3}\bigg(2m-\frac{\ga(a-2y+y^2)}{6(a-y^2)}\bigg)^2\,,\nonumber
\nd
densely defined on $L^2((y_-,y_+),\rho\,dy)$. We refer to the
aforementioned paper for more detailed discussions on the derivation of
the above formula\footnote{The approach taken in \cite{EK10} exploits the separability of the $\AdS^5\times
Y^{p,q}$ metrics to compute the eigenfunctions of the Laplace
operator in $Y^{p,q}$ in quasi closed form, by expressing them in
terms of the eigenfunctions of the Friedrichs extension of a single
second-order ordinary differential operator with four regular
singular points. The subtle geometry of the spaces $Y^{p,q}$
introduces additional complications in the analysis, since the
`angular' variables in which the metric of $Y^{p,q}$ separates are
not defined globally. In order to circumvent this problem the steps taken in \cite{EK10}
is to start
by constructing a Fourier-type decomposition of the space of
square-integrable functions on $Y^{p,q}$ adapted to the global
structure of the manifold and to the action of the Laplacian. Once
the eigenfunctions of the Laplacian in $Y^{p,q}$ have been computed,
the analysis of the Klein--Gordon equation in $\AdS^5\times Y^{p,q}$ can
be reduced to that of a family of linear hyperbolic equations in
anti-de Sitter space. In \cite{EK10} a detailed discussion of the existence and uniqueness of
causal propagators for these equations using Ishibashi and Wald's
spectral-theoretic approach to wave equations on static
space-times based on \cite{Wa80,IW03,IW04} were presented. Note that for our purpose, this presents
several  advantages over the classical method of Riesz transforms,
since the latter method only yields local solutions to the Cauchy
problem in the case in which the underlying space-time is not
globally hyperbolic~\cite{BGP07}.}.
We have set $\ga ~ \equiv ~ \si l/\tau$, and the function $v(\theta)$ defined in \eqref{scalarmode} satisfies the eigenvalue equation that comes from the
angular direction $\theta$ as:
\bg\label{evtheta}
\left[{1\over {\rm sin}~\theta} {\partial\over \partial\theta}{\rm sin}~\theta {\partial\over \partial\theta} -
{\left({n + 2m ~{\rm cos}~\theta\over {\rm sin}~\theta}\right)^2}\right]v_{nmj} = -\Lambda_{nmj} v_{nmj}\,.
\nd
The eigenvalues $\La_{nmj}$ are
given by the explicit formula:
\begin{equation}\label{eqval}
\La_{nmj}~ \equiv ~  2\Big[2j(j+1)+\big(|n+2m|+|n-2m|\big)(2j+1)+|n+2m||n-2m|+2m^2+n^2\Big]\,.
\end{equation}
In what follows we will drop the indices when there is no risk of confusion.

Before going on, it is worth recalling that the integers $n,m,l$ that
label the operators $S$ arise from the (quite subtle) Fourier
decomposition of functions we discussed in~\cite{EK10} and given above
in \eqref{scalarmode}, while the label $j$ (also an integer) was obtained by explicitly
solving an auxiliary eigenvalue problem associated with the geometry
of the sphere bundles (which had three regular singular
points). However, as we have already mentioned, there is little hope of
solving the eigenvalue problem for $S$ in closed form, since the spectral problem for the
operator $S$ is governed by a Heun differential equation. What we
will do, therefore, is to obtain some estimates for the eigenvalues of
$S$ that will allow us to approximate the conformal dimensions of the corresponding CFT.

\subsubsection{Behaviour of large eigenvalues (highly excited modes) 
\label{largek}}

In this subsubsection, we give an asymptotically exact result for
large energies (highly excited modes) of the operator $S\equiv S_{nmlj}$. The
basic idea is that, if we label the eigenfunctions of this operator by
an integer $k=1,2,\dots$, the $k$th eigenvalue is very close to a
constant multiple of $k^2$ for large $k$. To put it in a different
way, the eigenvalues tend to those of an infinite well, the width of
the well determined by the functions $P,Q,W$ that define the
Sturm--Liouville operator $S$. Very informally, the
justification would be that at high energies the leading terms are the
derivatives; this kind of asymptotic results are usually proved using
pseudo-differential operators.

The first observation is that, without any further assumptions, we
have an asymptotic formula (for large $k$, for highly excited modes) for the eigenvalues of $S$ namely:
the eigenvalues $\la_k\equiv \la_k(n,m,l,j)$ of $S$ are asymptotically given by following
expansion\footnote{A word of caution about the notation: The error term is $o(k^2)$, and not ${\cal O}(k^2)$.
The respective notations mean different things, and $o(k^2) \ll {\cal O}(k^2)$ for large $k$.
The notation $f(k)=o(g(k))$ means that $\lim_{k\to \infty} \frac{f(k)}{g(k)}=0$.
On the other hand the notation $f(k)= {\cal O}(g(k))$ means that that there exists a
positive constant $C$ such that for $k$ sufficiently large $\vert f(k)\vert \leq C \vert g(k)\vert$.
Simply, ${\cal O}(k^n)$ means that a term scaling like $k^n$ in the
proper limit of $k$, as familiar to physicists. Here $k \rightarrow
\infty$ is appropriate, and later in
\eqref{asymb} $a \rightarrow 0$ is so.
The two notions are different and in particular the $o(k^2)$ notation above
indicates that the error term grows slower than quadratically in $k$. If it had a power-law behaviour, it would be $o(k^2)={\cal O}(k^{2-\epsilon})$ with $\epsilon>0$. In some sense ${\cal O}$ is used when we know the power scaling of a term, and $o$ is when we know only the upper bound of the scaling.}
\begin{equation}\label{eq:1}
\la_k=C_0k^2+ {o}(k^2)\,,
\end{equation}
where the constant
\begin{equation}\label{cdefin}
C_0~ \equiv ~ 2\pi^2\bigg[\int_{y_-}^{y_+}\bigg(\frac{1-y}{a-3y^2+2y^3}\bigg)^{1/2}\,dy\bigg]^{-2}
\end{equation}
depends on the geometry of the manifold (that is, on $p$ and $q$)
through $y_\pm$ but
not on the Fourier modes $n,m,l,j$. (So that only the error term knows about these indices.)

The above statement is a consequence of general results in the theory of singular Sturm--Liouville operators. Indeed, it suffices to note that
$S$ is a lower-bounded one-dimensional self-adjoint operator, so it follows
from~\cite[Sec.\ 10.8]{Ze} that \eqref{eq:1} holds true with
\begin{equation}\label{cnow}
C_0~ \equiv ~ \bigg(\frac1\pi\int_{y_-}^{y_+}\big(w(y)\,r(y)\big)^{-1/2}\,dy\bigg)\,.
\end{equation}
An easy computation shows that this integral takes the above form, which can in turn be expressed in terms of elliptic functions.

Before ending this subsection, let us make the following remark.
Weyl's law\footnote{The Weyl law states that the first term in the asymptotic expansion for the $k$-th eigenvalue $\lambda_k$ of the Laplacian on an
$n$-dimensional compact Riemannian manifold is:
$$\lambda_k~ = ~ C_n k^{2/n}/({\rm Vol} ~M)^{2/n} +o(k^{2/n})$$
\noindent as $k \to \infty$. This was proved by Weyl in \cite{weyl1}. The second term was conjectured by Weyl in 1913 \cite{weyl2}
and proved only in 1980 by Ivrii \cite{ivrii}.}
ensures that, when the
eigenvalues of all the one-dimensional operators corresponding to the
various Fourier modes are taken into account, the eigenvalues of
the Laplace operator on $Y^{p,q}$ (let's call them $\widetilde\la_k$) obey the asymptotic
law
\begin{equation}\label{ldefin}
\widetilde\la_k= (2\pi)^2\bigg(\frac{5 k}{|\SS^4|{\rm Vol}(Y^{p,q})}\bigg)^{2/5}+{o}(k^{2/5})
\end{equation}
where $|\SS^4|$ denotes the volume of the unit 4-sphere and the volume of the manifold being given by~\cite{Sp04}
\begin{equation}\label{voldef}
{\rm Vol}(Y^{p,q})=\frac{\pi^2q^2[2p+(4p^2-3q^2)^{1/2}]}{3p^2[3q^2-2p^2+p(4p^2-3q^2)^{1/2}]}\,.
\end{equation}
Eq.~\eqref{eq:1} provides a somewhat more tangible way of presenting this asymptotic result in the sense that the asymptotics is separated into families labeled by additional ``quantum numbers''. A straightforward but tedious computation shows that, of course, when degeneracies are taken into account, the asymptotics~\eqref{eq:1} can be summed with respect to the additional ``quantum numbers'' to obtain \eqref{ldefin}.

Let us elaborate this a little bit more. We have seen that the analysis of the eigenvalues of the Laplacian in $Y^{p,q}$ can be
reduced to that of the eigenvalues of a family of one-dimensional operators $S=S_{nmlj}$. These operators are labeled by three integers $n,m,l$ and a
nonnegative integer $j$. Notice that if any of the quantum numbers $n,m$ or $l$ is nonzero (``higher Fourier modes''), all the eigenvalues of the
Laplacian corresponding to these quantum numbers are necessarily degenerate, as mapping $(n,m,l)$ to $(-n,-m,-l)$ leaves the eigenvalue equation
invariant. A convenient way of understanding the behavior of the eigenvalues if the Laplacian in geometric terms is the Weyl's law. For
this, let's denote by $\widetilde\lambda_k$ the $k$-th lowest eigenvalue of the Laplacian in $Y^{p,q}$, where each eigenvalue is repeated according to its
multiplicity. Obviously, for each $k$ there are ``quantum numbers'' $(n,m,l,j)$ such that $\widetilde\lambda_k= \lambda_{k'}(n,m,l,j)$ for some
$k'$.\footnote{It is worth emphasizing that one cannot explicitly compute the degeneracy of the eigenvalues, as there could be {\it non-geometric
degeneracies} in the sense that $\la_{k_0(n_0,m_0,l_0,j_0)}=\la_{k_1(n_1,m_1,l_1,j_1)}$ for some pair of indices {\em not} related by a symmetry of the
equation.} Weyl's law then ensures that the asymptotic distribution of the eigenvalues 
$\widetilde\lambda_k$ of the Laplacian is related to the volume of the
manifold through the relation \eqref{ldefin}.

\subsubsection{Bounds for the eigenvalues for small $a$ \label{smalla}}

In the previous subsubsection we obtained an asymptotic formula for the eigenvalues, which is asymptotically exact for large energies. It does not provide any information on low-lying eigenvalues, however. So our goal in this subsubsection is to provide some estimates for the whole spectrum in an appropriate regime. This regime will be the case when the parameter $a$ is small; as we will see, then we can obtain two-sided bounds for the
eigenvalues that provide an adequate control of the energies.

The technique we apply here is that, using the fact that $a$ is small $0<a<1$, we can Taylor expand the Laplacian operator in terms of small $a$ and drop higher orders of $a$ (as in \eqref{Snow}). Obviously this works the best if $a$ is very small, or equivalently when $q\ll p$,
but even moderately small $a$, it is a valid Taylor expansion. Instead of trying to obtain the spectrum of the original Laplacian operator, we use another operator \eqref{spec1} whose spectrum is exactly known as in \eqref{L.aux}. With an appropriate constant $C$ which does not depend on the parameters of the equation, we can compute the upper and lower bounds of the eigenvalues of Laplacian\footnote{An interesting question is how small $C$ can be, because if $C$ becomes large, the bound is very loose. Furthermore,
by comparing with the known low-lying scalar spectrum, we may learn something useful about $C$. The former and the latter points will be addressed in 
footnotes \ref{alberto1} and \ref{fang1} respectively.}.

Before passing to the actual derivation of the bounds, let us discuss the meaning of the smallness of $a$. It should be noticed that this is in fact a geometric hypothesis on the
manifold. In order to see this, let us recall the connection between
the parameter $a$ and the integers $p,q$ that controlled the geometry
of the bundle. In~\cite[Sec.\ 3]{Sp04} it is explained that the relationship between
$p,q$ and the endpoints $y_\pm$ is that
\begin{equation}
  \label{eq:2}
 y_+-y_-=\frac{3q}{2p}
\end{equation}
The idea now is that it can be easily seen that for any value of the
latter quotient we can find an $a$ for which~\eqref{eq:2} is
satisfied; indeed, $a$ can be chosen as
\bg\label{adefin}
a=\frac34\bigg[1-\frac{3q}{2p}-\bigg(1-\frac13\bigg(\frac{3q}{2p}\bigg)^2\bigg)^{1/2}\bigg]^2 -\frac14\bigg[1-\frac{3q}{2p}-\bigg(1-\frac13\bigg(\frac{3q}{2p}\bigg)^2\bigg)^{1/2}\bigg]^3
\nd
Hence it is not hard to see that $a\ll 1$ is equivalent to $q\ll p$,
so this condition translates immediately as a condition on the
geometry of the bundles. It this case,
\bg\label{anow}
a=\frac{27 q^2}{16 p^2}+\cO(q^3/p^3)
\nd
A closer look at the subsection on rational
roots in~\cite{Sp04} reveals that there is also an infinite number of
solutions with rational roots and arbitrarily small values of $a$
(recall that in this case the Sasaki--Einstein structure is
quasi-regular.)

The idea now is that, for very small $a$, the operator $-S$ should be
very similar to the one we obtain by dropping higher powers of $a$ (e.g.\ in the Taylor expansion of the coefficients),
namely
\bg\label{Snow}
-2\frac\pd{\pd y}(a-3y^2)\frac\pd{\pd
  y}+\frac{\ga^2}{2(a-y^2)}
+6\La+\frac{18(a-y^2)}{a-3y^2}\bigg(m+\frac{\ga y}{6(a-y^2)}\bigg)^2
\nd
This expression defines a self-adjoint operator on
$L^2(-(a/3)^{1/2},(a/3)^{1/2})$ via its Friedrichs extension (notice we still have too many singular points to solve the eigenvalue equation for $S$). It is
convenient to make things independent of $a$ by rescaling. For future
convenience, we introduce the variable $t\equiv a^{-1/2}y$ and, noticing
that
\bg\label{aredef}
\ga=\si l q(3a)^{1/2}\, (1+\cO(a))
\nd
we set $\bga \equiv a^{-1/2}\ga$ (observe that $\bar\ga$ still depends on
$a$, although it tends to a well-defined nonzero limit as $a\to0$). Here and in what follows, by $\cO(a)$
we will denote quantities bounded by a constant (independent of any
labels and of the geometry) times $a$, and whose derivatives satisfy
analogous bounds (i.e., behave like symbols with respect to these
bounds). We are thus led to consider (the
Friedrichs extension of) the operator
\bg\label{fredT}
T\equiv ~ -\frac\pd{\pd t}P(t)\frac\pd{\pd
  t}+Q(t)
\nd
in $L^2(I)$, with $I \equiv (-3^{-1/2}, 3^{-1/2})$ and
\
\begin{align*}
  P(t)& ~ \equiv~ 2 (1-3t^2)\,,\\
Q(t)& ~ \equiv ~ \frac{\bar\ga^2}{2(1-t^2)}
+6\La+\frac{18(1-t^2)}{1-3t^2}\bigg(m+\frac{\bar\ga t}{6(1-t^2)}\bigg)^2\,.
\end{align*}
In order to relate the spectral properties of $S$ (as an unbounded self-adjoint
operator on $L^2((y_-,y_+),\rho\, dy)$ to those of $T$ (on the space $L^2(I)$
with the standard Lebesgue measure $dt$), it
is convenient to start by relating these two $L^2$ spaces. An obvious
way to do so is through the following $a$-dependent change of variables:
\bg\label{adepch}
t ~\equiv ~ -\frac1{\sqrt3}+\frac2{\sqrt3}\frac{\int_{y_-}^y
  \rho(y')\,dy'}{\int_{y_-}^{y_+} \rho(y')\,dy'}~ \equiv ~ \mathcal T_a(y)\,.
\nd
This induces a unitary transformation $L^2((y_-,y_+),\rho\,dy)\to
L^2(I,dt)$, which transforms $S$ into the Sturm--Liouville operator of
the form:
\bg\label{slS}
{\widetilde S} \equiv ~ -\frac\pd{\pd t}{\widetilde P}(t)\frac\pd{\pd t}+ {\widetilde Q}(t)\,.
\nd
To derive the bounds, we start with the following observation:
the spectrum of the auxiliary operator
\bg\label{spec1}
T_{\mu}~\equiv~ -\frac\pd{\pd t}P(t)\frac\pd{\pd
  t}+\frac\mu{1-3t^2}
\nd
on $L^2(I)$, as a function of the parameter $\mu$, is given by
\bg\label{L.aux}
\ell_k(\mu)~ \equiv ~ \frac{3}2\bigg(1+\sqrt{\frac{8\mu}{3}}+2k\bigg)^2-\frac{3}2
\nd
The proof of the above statement can be argued using
a straightforward computation. To start, observe that it suffices to see that the exponents
of the equation $T_\mu f=-\ell f$ are
\bg\label{expos}
\pm \sqrt{\mu\over 24}, ~~~~~~~~~~ {1\over 2}\left(1\pm \sqrt{1+{2\la\over 3}}\right)
\nd
at (0 and at 1) and at $\infty$ respectively.
The eigenvalues
then arise as the necessary condition for
\bg\label{eigval}
(1-3t^2)^{-(\mu/24)^{1/2}} f(t)
\nd
to be a polynomial in $t$, thus proving the required statement.

After developing the necessary mathematical preliminaries,
we are now ready to compute bounds for the eigenvalues of $S$ (which
coincide with those of $\tS$, by definition). Notice that we cannot
obtain bounds using a relative compactness argument, as any
perturbation of the function $P$ will lead to
corrections that are not relatively compact with respect to the
original operator (because they have the same number of
derivatives as the initial operator). What we can do is to exploit monotonicity using the
following two observations.
The first observation is that
there is a constant $C$, which does not depend on the parameters
  of the equation, such that the following bounds for ${\widetilde P}(t)$ hold for all $t\in I$:
\bg\label{horibol}
(1-Ca)P(t)\leq {\widetilde P}(t)\leq (1+Ca)P(t)\,.
\nd
This inequality is obvious in view of the formula \eqref{adepch} for the map
$y\mapsto t$, and simply asserts (roughly speaking) that the map does not alter
the singularities too much.

Our second observation is somewhat similar to the first one in the sense that we again claim that
there is a constant $C$, which does not depend on the parameters
  of the equation, such that the following bounds for ${\widetilde Q}(t)$ hold for all $t\in I$:
\bg\label{qbound}
&&{\widetilde Q}(t)\geq(1-Ca)\bigg(\frac{\mu_-}{1-3t^2}+\frac{1+\bga^2}2+6\La-Ca\bigg)\,,\nonumber\\
&&{\widetilde Q}(t)\leq (1+Ca)\bigg(\frac{\mu_+}{1-3t^2}+\frac{3(1+\bga^2)}4+6\La+Ca\bigg)\,,
\nd
where $\mu_+$ and $\mu_-$ are defined in the following way:
\bg\label{mupmdef}
\mu_-~ \equiv ~ 12\max\bigg\{0, m-\frac{\bga}{4\sqrt3}\bigg\}^2\,,\qquad \mu_+:~ \equiv ~ 18\bigg(m+\frac{\bga}{4\sqrt3}\bigg)^2\,.
\nd
The proof of the above two inequalities
are a straightforward consequence of
the fact that
\bg\label{conseq}
(1-Ca)(Q(t)-Ca)\leq {\widetilde Q}(t)\leq (1+Ca)(Q(t)+Ca)\,.
\nd
(One might wonder
why we included an additive error $Ca$ here and not in the
estimate for $\tP$. The reason is that $\tP$ does not vanish in the interval $I$,
and this is enough for us to control the error via a multiplicative
constant.)

It is standard that if we take nicely behaved functions $P_j(t)$ and $Q_j(t)$ on $I$, with $j=1,2$ and $P_j(t)>0$, and suppose that $P_1(t)\geq P_2(t)$ and $Q_1(t)\geq Q_2(t)$ (resp.\ $P_1(t)\leq P_2(t)$ and $Q_1(t)\leq Q_2(t)$), then the $k$-th eigenvalue of (the Friedrichs extension of) the operator $-\frac d{dt}P_1(t)\frac d{dt}+Q_1(t)$ is larger or equal (resp.\ smaller or equal) than those of  $-\frac d{dt}P_2(t)\frac d{dt}+Q_2(t)$. Hence it is elementary to derive the bounds
\bg\label{sbound}
\Lambda_k^{[-]} ~ \leq ~\lambda_k ~ \leq ~\Lambda_k^{[+]}
\nd
where
\bg\label{sbound2}
&& \Lambda_k^{[-]} ~ = ~ (1-Ca)\bigg(\ell_k(\mu_-) +\frac{1+\bga^2}2+6\La-Ca\bigg)~, \nonumber\\
&& \Lambda_k^{[+]} ~ = ~ (1+Ca)\bigg(\ell_k(\mu_+) +\frac{3(1+\bga^2)}4+6\La+Ca\bigg)~,
\nd
from the inequalities \eqref{horibol} and \eqref{qbound}, the formula for the eigenvalues $\ell_k(\mu)$ of the auxiliary operator $T_\mu$  derived in \eqref{L.aux} and elementary inequalities in $I$ such as
\bg\label{elemin}
1\leq  (1-t^2)^{-1}\leq
3/2\,.
\nd
The bounds \eqref{sbound}, in which $C$ stands for an $a$-independent constant and $\ell_k(\mu)$ is given by \eqref{L.aux}, constitute the main result of 
this subsection\footnote{One might worry about the strength of our bound. For example a question would be whether the bound could be loose if 
constant such as $C$ is large. 
To answer this we first note that 
the constants do not arise exactly from a power series
expansion, but rather as the Taylor formula with estimates for the
remainder (which is essentially the mean value theorem). Therefore,
the constant $C$ can be explicitly computed as the (sum of the)
supremum (for $t$ and $a$ between certain values) of the derivative of
some functions appearing in $P$ or $Q$ with respect to the parameter
$a$. For this reason, the behavior of this constant is controlled, and
can be computed explicitly. For example, a rough computation reveals that the
constant $C$ can be chosen to be of order 10 when $a$ is smaller than
$10^{-n-1}$, so the relative error is at most of order $10^{-n}$. Since $a = (q/p)^2$ upto
some inessential factors, it is enough that $q/p < 10^{-(n+1)/2}$.
These estimates can be refined easily. \label{alberto1}}.

As a remark, notice that the above bounds also ensure that the eigenvalues have the asymptotic behavior
\bg\label{asymb}
\la_k=6(1+\cO(a))k^2+\cO(k)\,.
\nd
This is precisely the growth rate computed in \eqref{eq:1},
since it is easy to see that the constant
\bg\label{conow}
C_0~ \equiv
{2\pi^2}
  \bigg[\int_{y_-}^{y_+}\bigg(\frac{1-y}{a-3y^2+2y^3}\bigg)^{1/2}\,dy\bigg]^{-2}
\nd
entering Weyl's law \eqref{cdefin}  tends to $6$
as the constant $a$ tends to $0$.

\section{Examples of scalar and other modes \label{example}}

Now that we have discussed the spectrum of scalar modes in the internal $Y^{p,q}$ space, it is time to study some examples of these
modes. However before moving ahead we should point out that
in this section (and also the next) we will {\it not} address the spectra of the theory. To analyse the spectra (for example along the lines 
of \cite{jatkar, ceresole1, ceresole2}) 
we would not only need to go beyond the 
scalar fields, but would also require exact eigenvalues of the KK modes for all spin-states of the theory $-$ a calculation that will be relegated for future works. 
The advances that we made in the previous section
is a good starting point and we will benefit from further development. 
At this point we will suffice ourselves by studying some basics aspects of scalar and other modes 
from supergravity perspective in this section. In the 
next section we will discuss possible non-conformal extensions of our model. Again the emphasis therein would be to study the supergravity background and not the 
matching of spectra. 
 
The simplest examples of scalar and other modes 
that appear for our case are from the decomposition of the 2-forms in \eqref{decomp2}. These decompositions lead to
two possible theories on the boundary where we define the CFTs.

\vskip.1in

\noindent $\bullet$ Non-commutative geometry: Let us consider the NS $B$ field with both components along the boundary, i.e we can switch on $B_{ij}(x)$
where $i, j = 1, 2, 3$ and $x_\mu$ specify coordinates in $AdS_5$ space, leading to non-commutative geometry in the dual gauge theory.
For example, a $B$-field component of the form $B_{ij}(r)$, with
$r$ being the radial direction in the $AdS_5$ space, would be able to generate non-commutative theory on the boundary.
Clearly this mode is a scalar mode in the internal $Y^{p, q}$
space.

\vskip.1in

\noindent $\bullet$ Dipole theory: This time we consider
the NS $B$-field which has one component along the boundary and the other component either along the radial $r$
direction or along the internal $Y^{p, q}$ directions. Consider first a component of the NS $B$ field of the form $B_{ir}$. However if this
component is only a function of $x^\mu$, then we can make a gauge transformation to rotate the NS $B$ field components
along the boundary which in turn will convert the boundary theory to a non-commutative theory. The other alternative is to
make it
gauge equivalent to zero for the $B$ field component of the form $B_{ir}(r)$. Thus the only non-trivial cases appear to be of the form
$B_{ir}(y), B_{ia}(x, y)$ and they both lead to the dipole theories. However none of these are scalar modes in the internal $Y^{p, q}$.
The special case where the NS $B$ field is of the form $B_{ia}(x, y)$
fits in with our decomposition \eqref{decomp2}, and leads to a simple vector decomposition of the boundary theory.

Thus the simplest scalar mode leading to
noncommutativity can be specified by a 2-form $\theta^{ij}$
such that the commutator of the coordinates
on the boundary theory is $[x^i, x^j]  = i\theta^{ij}$.
The parameter $\theta^{ij}$ has dimensions $-2$.
At low-energies, noncommutative super Yang-Mills theory (NCSYM) can be described by augmenting the action with:
\bg\label{ncgact}
\int\theta^{ij}{\cal O}_{ij}(x) d^4 x,
\nd
where ${\cal O}_{ij}$ is an operator of dimension $6$ in the superconformal
SYM on a commutative space.
In the conventions such that the SYM Lagrangian is:
\bg\label{syma}
{\cal L}_{\rm SYM} = {\rm tr}\left[
\frac{1}{2g^2}\sum_{I=1}^6\partial_{i}\phi^I\partial^{i}\phi^I
+\frac{1}{4g^2}F_{ij}F^{ij}
  +\frac{1}{2g^2}\sum_{I<J}\left[{\phi^I}, {\phi^J}\right]^2\right]
  + {\rm fermions},
\nd
the bosonic part of the operator ${\cal O}_{ij}$ can be written as:
\bg\label{bospart}
{\rm tr}\left[
\frac{1}{2g^2}F_{jk}F^{kl}F_{li}
-\frac{1}{2g^2}F_{ij}F^{kl}F_{kl}
+\frac{1}{g^2}F_{ik}\sum_{I=1}^6\partial_{j}\phi^I\partial^{k}\phi^I
-\frac{1}{4g^2}F_{ij}\sum_{I=1}^6\partial_{k}\phi^I\partial_{k}\phi^I\right].\nonumber\\
\nd
Here, $g$ is the SYM coupling constant, $F_{ij}$ is the $U(N)$
field-strength, and $\phi^I$ ($I=1\dots 6$) are the scalars.

For the second case we expect the boundary theory to be deformed
by an operator of the form ${\cal O}_i$. The deformation by $L^i\cO_i$ (where
$L^i$ is a constant vector) is the low-energy expansion of
a nonlocal field-theory, the so-called dipole-theory, described in \cite{BG, DGR, BDGKR}.

Furthermore, as discussed in \cite{BG} (see also \cite{DGR, BDGKR, DJ}),
the bosonic part of the SYM operator ${\cal O}_i$
can be calculated by changing to local variables (see \cite{BG}
for more details).
We can write it in ${\cal N} = 1$ superfield notation as \cite{BG}:
\bg\label{difuli}
{\cal O}_i =
  {i\over {g_{YM}^2}}\int d^2\theta \epsilon^{ab}
   {\rm tr}\left[\sigma_i^{\alpha\dot \alpha} W_\alpha \Phi_a D_{\dot \alpha}\Phi_b +\Phi \Phi_a D_i\Phi_b\right]
   + {\rm c.c.}
\nd
Here, we denote the
${\cal N}= 1$ chiral field as $\Phi$ and the ${\cal N} = 1$
vector-multiplet with the field-strength $W_\alpha$. The original ${\cal N} = 2$ hypermultiplet is now written in terms of the two
${\cal N} = 1$ chiral multiplets
$\Phi_a$ ($a=1,2$). Finally, $\sigma_i^{\alpha \dot \alpha}$ are Pauli matrices. As expected, the operator ${\cal O}_i$ has conformal dimension 5.

\subsection{Possible type IIA brane realisation \label{IIAbrane}}

In the following we will discuss these backgrounds in somewhat
more details by switching on appropriate $B$ fields. This is slightly different from
allowing the $B$ field as a {\it fluctuation}. A non-trivial background $B$ field will change the geometry in some particular way which would
reflect the corresponding backreactions. To analyse the corresponding backreactions we have to study the scenario directly from $N$ D3-branes
probing the geometry given by a cone over the $Y^{p, q}$ spaces. This starting point in fact has many intriguing possibilities in addition to the
ones related to generating non-local field theories. One of the possibilities is to see whether a brane realisation of the form \cite{Dasgupta:1998su}
in type IIA can also be made for our case. We will therefore start by analysing this interesting possibility
first and then go for the non-local theories.

To study D3-branes at the tip of a cone over the $Y^{p, q}$ manifolds, we will assume the usual ansatz for the D3-brane metric given in terms of a
harmonic function $H$ which is typically a function of $r$ and the $Y^{p, q}$ coordinates. Let us therefore
take the following metric ansatz:
\begin{eqnarray}\label{mota1}
ds_{\rm IIB}^2=H^{-1/2}ds_{0123}^2+H^{1/2}(dr^2+r^2dM_5^2),
\end{eqnarray}
where $dM_5^2$ is the same in eq. \eqref{stdm} and
$F_5=(1+\ast)d\beta_0 \wedge dx_0\wedge dx_1 \wedge dx_2 \wedge dx_3$
with $H=1+\frac{r_0^4}{r^4} \equiv \beta_0$.
We also assume the dilaton is
zero. As in \cite{Dasgupta:1998su}, the internal metric has three
isometries along the $\alpha$, $\psi$ and $\phi$ directions. We first
do a T-duality along $\alpha$ direction. The metric becomes
\begin{eqnarray}\label{tuamet}
ds_{\rm IIA}^2&=&H^{-\frac{1}{2}}ds_{0123}^2+H^{\frac{1}{2}}\Bigg\{dr^2+r^2\Big[\frac{1-cy}{6}(d\theta^2+\sin^2\theta
d\phi^2)+\frac{1-cy}{2f(y)}dy^2\nonumber\\
&&+\frac{f(y)}{9(a-y^2)}(d\psi^2-\cos\theta
d\phi)^2+\frac{(1-cy)}{2H r^4(a-y^2)}d\alpha^2\Big]\Bigg\}\nonumber\\
 & = & H^{-\frac{1}{2}}\left[dx_{0123}^2 + \frac{1-cy}{2r^2(a-y^2)} d\alpha^2\right] + H^{\frac{1}{2}}\Bigg\{dr^2+r^2\Big[\frac{1-cy}{6}(d\theta^2+\sin^2\theta
d\phi)^2\nonumber\\
&& + \frac{1-cy}{2f(y)}dy^2 +\frac{f(y)}{9(a-y^2)}(d\psi-\cos\theta
d\phi)^2\Big]\Bigg\},
\end{eqnarray}
with the following two components of the $B$-fields:
\begin{eqnarray}
B_{\alpha\psi}=\frac{ac-2y+y^2c}{6(a-y^2)},\quad\quad
B_{\alpha\phi}=-\frac{ac-2y+y^2c}{6(a-y^2)}\cos\theta,
\end{eqnarray}
and the original D3 branes
become D4 branes. The existence of the two $B$-fields might indicate the possibility of two NS5 branes, provided $H_{\rm NS} = dB$ is a source term
and the integral of $H_{\rm NS}$ over a three-cycle is an integer. The first one is harder to determine because 
the knowledge of the global behavior of the two $B$-field components is lacking, although the metric that we are dealing with 
is global.
This is because we {\it delocalized} along the $\alpha$
direction to make the harmonic function $H$ independent of that direction so that T-duality rules of \cite{BHO} could be implemented. This is of course
a slight oversimplification as this works well for some purposes, but not others.
The harmonic function should be taken to be a function of $\alpha$ as well, and then one may T-dualise the
background using the technique illustrated in \cite{GHM}. Under such a T-duality both the $B$-field components will pick up dependences on $H$ as well.
We will discuss more on this a little later.

For the second case, one may do
better by converting the three-forms to two-forms and integrating over two-cycles. This can be easily achieved by making a U-duality transformation
of the form $T_\alpha ST_3$ where $S$ denotes a S-duality transformation and $T_m$ denotes a T-duality along $x^m$ direction.
Thus making a T-duality along $x_3$ direction we get the following metric in type IIB theory:
\begin{eqnarray}
ds^2&=&H^{-\frac{1}{2}}dx_{012}^2+H^{\frac{1}{2}}\Bigg\{dx_3^2+dr^2+r^2\Big[\frac{1-cy}{6}(d\theta^2+\sin^2\theta
d\phi^2)+\frac{1-cy}{2f(y)}dy^2\nonumber\\
&&+\frac{f(y)}{9(a-y^2)}(d\psi^2-\cos\theta
d\phi)^2+\frac{(1-cy)}{2Hr^4(a-y^2)}d\alpha^2\Big]\Bigg\}.
\end{eqnarray}
Under this T-duality the D4 branes become D3 branes but extending along $x^0$, $x^1$, $x^2$ and
$\alpha$ directions. However the $B$-fields remain unchanged. If these $B$-fields are coming from some source NS5-branes, then the NS5-branes
would not change under the T-duality.

Let us now do the S-duality under which the NS $B$-fields become RR $B$-fields
and the metric gets an overall factor from the dilaton field
$\sqrt{\frac{2r^2(a-y^2)}{1-cy}}$ while the D3 branes remain the same. When we T-dualise this background along $\alpha$ direction, the metric becomes
\begin{eqnarray}
ds^2&=& {\cal H}^{-\frac{1}{2}}dx_{012}^2+H {\cal H}^{-\frac{1}{2}}\Bigg\{dx_3^2+dr^2+r^2\Big[\frac{1-cy}{6}(d\theta^2+\sin^2\theta
d\phi^2)+\frac{1-cy}{2f(y)}dy^2\nonumber\\
&&+\frac{f(y)}{9(a-y^2)}(d\psi^2-\cos\theta
d\phi)^2+ d\alpha^2\Big]\Bigg\},
\end{eqnarray}
and the RR three-form fields become the type IIA gauge fields. We have also defined ${\cal H} = {H\over 2r^2} \frac{1-cy}{a-y^2}$ as our modified
harmonic function.
If these gauge fields are sourced by D6 branes then they are the ones that come from the
type IIB D5 branes. The D3 branes on the other hand become D2 branes. Lifting this configuration to M-theory the eleventh direction has the required
local ALE fibration with M2 branes at a point on the four-fold.

The above set of manipulation is suggestive of NS5 branes in the original type IIA configuration provided the gauge field EOM has a source term. Thus
if we write the local type IIA gauge field over a patch as:
\bg\label{iiagf}
A = A_\psi d\psi + A_\phi d\phi \equiv \frac{ac-2y+y^2c}{6a-6y^2} \Big[{\cal F}_1(H) d\psi - {\cal F}_2(H) {\rm cos}~\theta ~d\phi\Big],
\nd
where we have inserted the correction from the harmonic function as ${\cal F}_{1,2}(H)$,
then there exists a global field strength $F = dA$. Now if it satisfies the two conditions mentioned earlier, namely
\bg\label{chukka}
d\ast F = {\rm sources}, ~~~~~~ \int_{S^2} F = {\rm integer},
\nd
then this would not only help us to identify the NS5 branes in the original type IIA set-up, but also help us to {\it count} the number of the
NS5 branes.
Such a source term in \eqref{chukka}
may not be too difficult to see from our analysis if we take \eqref{iiagf} seriously. The LHS of \eqref{chukka} will
involve terms like $d\ast d{\cal F}_1(H)$ and $d\ast d{\cal F}_2(H)$. Since
$\square H = \ast d\ast d H$ lead to source terms in the supergravity solution, it should be
no surprise if the above two terms in \eqref{chukka} coming from ${\cal F}_{1,2}(H)$ lead to D6 brane source terms in our model.

The above analysis is definitely suggestive of this scenario, although the precise orientations of the NS5 branes are not clear to us
at this stage. Furthermore there is the subtlety pointed out in \cite{royston} which we might have to consider too.
Note also that from \eqref{tuamet} the D4 branes are wrapped along a non-trivial $S^1_\alpha$ cycle.
More details on this will be relegated to future works.

Before we end this subsection, we would like to point out another scenario related to the type IIB metric \eqref{canm}. As has been described
earlier, \eqref{canm} is related to \eqref{stdm} by a series of coordinate transformations. Interestingly the metric \eqref{canm} is
closely related to the conifold metric if one makes the following substitutions in \eqref{canm}:
\bg\label{substit}
c = 0, ~~~~~~ a = 3, ~~~~~~ y = -{\rm cos}~\theta_2, ~~~~~~ \beta = \phi_2, ~~~~~~ \theta = \theta_1, ~~~~~~\phi = \phi_1
\nd 
where $\beta$ was defined in \eqref{newcoor}.
So a natural question to ask would be what happens if one makes a T-duality along the $\psi$ direction. It is of course well known that, in the
limit \eqref{substit}, a T-duality along $\psi$ direction leads to an orthogonal (not necessarily intersecting) NS5 branes configuration \cite{Dasgupta:1998su}.
If we now make a T-duality along $\psi$ direction,
the metric that we get in type IIA side is the following:
\begin{eqnarray}\label{mota2}
ds^2&=&H^{-1/2}\left[dx_{0123}^2+\frac{18(1-cy)}{r^2W}d\psi^2\right]+H^{1/2}\Bigg[dr^2+r^2\Big(\frac{1-cy}{6}(d\theta^2+\sin^2\theta
d\phi^2)\nonumber\\
&&\quad\quad\quad\quad\quad\quad\quad\quad\quad\quad\quad\quad\quad\quad\quad\quad\quad\quad+\frac{1-cy}{2f}dy^2+\frac{4f}{W}d\alpha^2\Big)\Bigg],
\end{eqnarray}
where $W=3c^2y^2-6cy+2+ac^2$.
Interestingly, we find the metric has the simpler form without cross-terms at all. This is again reminiscent of \cite{Dasgupta:1998su}. We also
find two NS $B$ fields whose components are given as:
\begin{eqnarray}\label{mota3}
B_{\psi\alpha}=\frac{6(ac-2y+cy^2)}{W}, \quad \quad
B_{\psi\phi}=-\cos\theta.
\end{eqnarray}
The absence of a cross-term is not a big surprise because we can rewrite \eqref{canm} in a suggestive way using the coordinates \eqref{substit}
and taking ($c, a$) away from the conifold value ($0, 3$). The metric \eqref{canm} becomes:
\bg\label{canmbec}
ds^2 &=& a_1(d\theta_1^2 + {\rm sin}^2~\theta_1 d\phi_1^2) + \Big[a_2 ~{\rm sin}^2~\theta_2 d\theta_2^2 + a_3~(d\phi_2
+ c~{\rm cos}~\theta_1 d\phi_1)^2\Big]\nonumber\\
&& ~~~~~~~~~~~ + {1\over 9}\Big[d\psi + (1 + c~{\rm cos}~\theta_2) {\rm cos}~\theta_1 d\phi_1 - {\rm cos}~\theta_2 d\phi_2\Big]^2,
\nd
where $a_1, a_2$ and $a_3$ are given by the following expressions:
\bg\label{a1a2a3}
a_1 = \frac{1 + c~{\rm cos}~\theta_2}{6}, ~~~ a_2 = \frac{1}{2}\cdot 
\frac{1 + c~{\rm cos}~\theta_2}{a - 3{\rm cos}^2\theta_2 - 2c~ {\rm cos}^3\theta_2},
~~~a_3 = \frac{1}{18}\cdot\frac{a - 3{\rm cos}^2\theta_2 - 2c {\rm cos}^3\theta_2}{1 + c~{\rm cos}~\theta_2}.\nonumber\\
\nd
A T-duality along $\psi$ direction will give us the configuration that we discussed above
(using non-canonical coordinates)\footnote{Note however that \eqref{mota2} and the T-dual of \eqref{canmbec} may look different because in \eqref{mota2}
one cannot substitute the coordinate transformation directly as the coordinates of \eqref{mota2} are the T-dual coordinates of \eqref{stdm}. Thus
a simple substitution of $\alpha = -{1\over 6} (\phi_2 + c\psi)$ in \eqref{mota2} cannot be done.}.
To see what \eqref{mota2} and \eqref{mota3} imply, let us again go to the limit where
$c=0$ and $a=3$. In this limit\footnote{For all other purposes we set $c = 1$.}
we recover the exact brane picture of type IIA
discussed in \cite{Dasgupta:1998su}.
This may mean that we have some NS5 branes along the ($\theta$, $\phi$)
directions and some NS5 branes along ($\alpha$, $y$) directions (or in a more canonical language, we have a set of NS5 branes along ($\theta_1, \phi_1$)
directions and another set of NS5 branes along ($\theta_2, \phi_2$) directions).
These two set of NS5 branes are locally orthogonal to each other, so as to
preserve ${\cal N} = 1$ supersymmetry. The $d\psi$ fibration structure in \eqref{canmbec} also tells us that there are two local $B$-fields in type IIA
side that would
T-dualise to give us the required background \eqref{canmbec}.
The $N$ type IIB D3-branes become $N$
of D4 branes along $\psi$ direction suspended between these NS5
branes.

Unfortunately the $c \ne 0$ scenario is not quite the same as the simpler ($c, a$) $=$ ($0, 3$) scenario. In
particular\footnote{We will henceforth use only the non-canonical coordinates by choice. An equivalent construction could be easily done with the
canonical coordinates \eqref{substit}.} at $y=y_1$ and $y=y_2$ the metric \eqref{mota2} develops conical singularities, in other words
now $y$ and $\alpha$ no longer form a sphere. This can be easily seen by
taking the limit $y\rightarrow y_i$ where $i=1,2$. In this limit we
can write the metric along the $y$ and $\alpha$ directions as:
\begin{eqnarray}\label{2sphe}
\frac{1-y_i}{f'_i(y-y_i)}dy^2 ~+ ~ \frac{4f'_i(y-y_i)}{W_i}d\alpha^2.
\end{eqnarray}
This is not quite the metric of a 2-sphere. To see this more clearly, let us define a quantity $R$ in the following way:
\bg\label{rdef}
R ~ \equiv~ 2\sqrt{(1-y_i)(x-y_i)\over f'_i}.
\nd
Using this defination we can rewrite the metric \eqref{2sphe} in a bit more suggestive way:
\begin{eqnarray}\label{2sphe2}
dR^2 ~+ ~ \frac{f_i^{'2}R^2}{(1-y_i)W_i}d\alpha^2.
\end{eqnarray}
Clearly the above metric becomes the metric of a 2-sphere 
only when $\alpha$ is periodic with a period of $L \equiv 2\pi
\sqrt{(1-x_i)W_i}/f'_i$. However recall that instead $\alpha$ has a
period of $l\neq L$. This means we will always have two conical
singularities at $ y = y_i$.

Let us now prove that there are no other singularities in this metric.
Notice that other singularities can happen only at $W=0$, which has
two roots:
\bg\label{2roots}
y_{\pm} ~ = ~ 1\pm \sqrt{\frac{1-a}{3}}.
\nd
Since $y_+ > 1$, it is clear
that $y_+$ is already out of the range of $y$,  while it is not so
obvious for $0 < y_- < 1$. To see the range of $y_-$, we substitute $y_-$ into $f$ to get:
\begin{eqnarray}
f(y_-)~ = ~ -\frac{2}{3\sqrt{3}}(1-a)\sqrt{1-a} ~ < ~ 0,
\end{eqnarray}
which means $y_-> y_2$ and therefore it is also out of the range. Therefore there are no other singularities in this metric. 

The above picture gives us an indication how the brane dual could be constructed although the actual details are much harder to present than our
previous construction. It is also true that
the delocalization effects are again present in the harmonic function but this time, thanks to the
canonical representation of the metric \eqref{canmbec}, a direct mapping to the intersecting brane configuration for $c=0, a=3$ gives us a hope that
similar brane dual description does exist for generic cases (although at this stage one may need to consider the subtleties pointed out in
\cite{royston}).
The interesting thing however is that a T-duality along $\alpha$ also seems to
lead to a similar configuration provided of course \eqref{chukka} holds. This shouldn't be a surprise because $\alpha$ and $\psi$ are related by a
linear coordinate transformation for $c \ne 0$.

\subsection{Non-commutative and dipole deformations \label{dipole}}

The above T-duality arguments give us a way to study the underlying ${\cal N} = 1$ gauge theory from two different point of views: one directly from
$N$ D3 branes at the tip of the cone in type IIB theory,
and other from $N$ D4 branes in a configuration of two orthogonal set of NS5 branes in type IIA theory; although for the latter case the
precise orientations of the two NS5 branes still need to be determined.

The non-commutative and the dipole deformations could also be studied from these two viewpoints. However in this paper we will not consider the
type IIA brane interpretations of these deformations. Here we will suffice with only the type IIB description and a fuller picture will be elaborated
in a forthcoming work.

Our starting point is the well known observation that
once we have a solution we can use $TsT$ to deform it into various
different solutions, where $T$ is a T-duality transformation and $s$ is a shift.

Given the background metric \eqref{mota1} with D3 branes we have three
kinds of deformations:

\vskip.1in

\noindent $\bullet$ T-dualise along one space
direction say $x_3$ then shift along another space direction say $x_2$ mixing ($x_2, x_3$)
and then T-dualise back along $x_3$ direction.

\vskip.1in

\noindent $\bullet$ T-dualise along $x_3$ and then shift\footnote{Again mixing $x_3$ with one of the internal directions.}
along one of the
internal directions that are isometries of the background namely along $\alpha, \phi$ or $\psi$
and then T-dualise back along $x_3$ direction.

\vskip.1in

\noindent $\bullet$ T-dualise, shift and then T-dualise along internal
directions.

\vskip.1in

\noindent The first of these operations would lead to the non-commutative gauge theory on the D3 branes
whose details we discussed earlier. For the other two cases, the set of operations may lead to
non-local dipole theories on the D3 branes.

In this paper we only study the first kind of deformation, whose advantage is that the
internal metric remains unchanged so our scalar modes analysis in
$Y^{p,q}$ is still valid. Of course this still doesn't help us to get the exact matching of spectra as we pointed out earlier. Therefore, in the following, we will 
briefly spell out the supergravity background. 
For the rest two kinds of deformations our
analysis generally cannot be applied as the internal metric will change quite a bit. We will leave a detailed analysis of
dipole deformations for future works.

For the non-commutative case, the starting point would be the choice of the shift after a T-duality along the $x_3$ direction.
We choose the shift to be
\begin{eqnarray}
x_2\mapsto \frac{x_2}{\cos\theta}+x_3~ \sin\theta,\quad x_3\mapsto
x_3 ~\cos\theta.
\end{eqnarray}
After the series of duality transformations the background can be easily determined to take the following form:
\begin{eqnarray}\label{ncg}
ds^2=\frac{1}{\sqrt{H}}\Big[-dx_0^2+dx_1^2+J(dx_2^2+dx_3^2)\Big]+\sqrt{H}(dr^2+d\mathcal{M}^2_5),
\end{eqnarray}
which clearly tells us that the internal $Y^{p, q}$ space do not change, but the Lorentz invariance along the $x^2$ and $x^3$ direction is
broken as one would have expected. The metric has the same form as in
\cite{Maldacena:1999mh} and the gauge theory on D3 branes should be
non-commutative in $x_2$ and $x_3$ directions. The non-commutativity parameter, which is the $B_{23}$ field, and the Lorentz breaking term $J$, in
\eqref{ncg}, are defined in the following way:
\bg\label{dhonsho}
J~ =~ {H \over \sin^2\theta + H \cos^2\theta}, ~~~~~~ B_{23} ~= ~ {\tan\theta \over \sin^2\theta + H \cos^2\theta}.
\nd
This completes our discussion of the conformal models related to the $Y^{p, q}$ spaces. In the following section we will discuss the
non-conformal extensions of the above models. We will specifically concentrate on the possibility of geometric transitions in these models.

\section{Non-conformal duals and geometric transitions \label{geomtr}}

\begin{figure}[htb]\label{dualitiesGT}
        \begin{center}
\includegraphics[height=10cm]{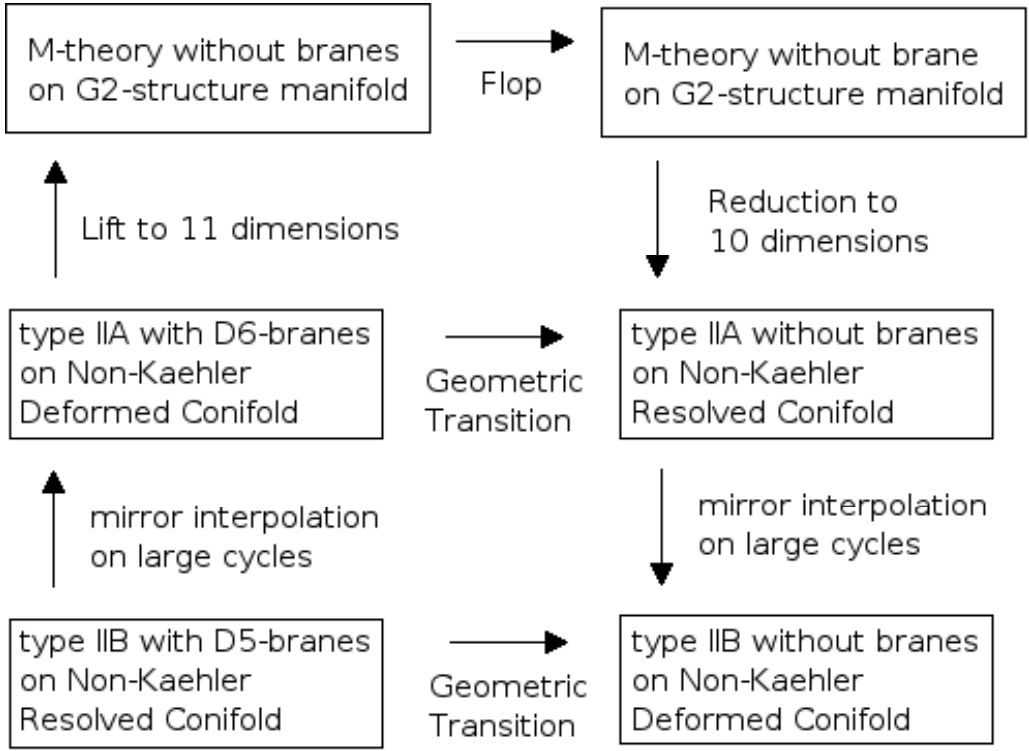}
        \caption{The duality map to generate the full geometric transitions in the supersymmetric global
set-up of type IIA and type IIB theories.}
        \end{center}
        \end{figure}

The non-conformal duals to the $Y^{p, q}$ spaces, along the lines of the cascading model of \cite{KS}, have already been addressed in the literature (see
for example \cite{klebher, cotrone2} etc). The UV gauge groups for $Y^{p, 1}$ and $Y^{p, p-1}$ are respectively given in equations (75) and (87) of
\cite{klebher}. For both the cases the IR gauge group is:
\bg\label{farIR}
SU(M) \times SU(2M) \times ... \times SU(2pM)
\nd
where $M$ denotes the number of D5 branes wrapping the two-cycles of $Y^{p, 1}$ and $Y^{p, p-1}$ spaces. Such a gauge group is more complicated than the
simple picture that we had for \cite{KS} and therefore the {\it far} IR picture could be more involved: there could be non-trivial baryonic branches.
This story has not yet been fully clarified, and therefore it gives hope that the brane picture that we developed here may help us to study the
far IR picture in more
details\footnote{For example the brane picture developed for the $T^{1, 1}$ case in \cite{dot} clearly showed how the far IR physics for
cascading theory could be understood. We expect similar story to unfold here too.}.
We will however not pursue the cascading story anymore here. Instead we go to a slightly different direction that
may provide us with an alternative way to study the far IR physics of these models \cite{pandoman, chen1}.

Our starting point would be to ask whether the far IR physics of the non-conformal set-up could be likened to the geometric transtion 
story \cite{vafaGT} that we
developed in the series of papers starting with \cite{gtpaper} and culminating with \cite{chen1}.
For the geometric transition picture to hold, we need few essential
ingredients:
\vskip.1in

\noindent $\bullet$ Resolution and deformation for the cone over $Y^{p, q}$. These resolved and deformed spaces are not required to be
Calabi-Yau spaces, but they should have at least $SU(3)$ structures (in the presence of branes and fluxes)
so that supersymmetric models could be constructed.

\vskip.1in

\noindent $\bullet$ Supersymmetric configurations with D5 branes wrapped on two-cyles of the resolved $Y^{p, q}$ and D6 branes wrapped on
three-cycles of the deformed $Y^{p, q}$ including supersymmetric
configurations {\it without} branes but with fluxes. Again the {\it overall} pictures for both
cases should preserve $SU(3)$ structures.

\vskip.1in

\noindent $\bullet$ Two kinds of $G_2$ structure manifolds should exist in M-theory. One, the lift of the deformed $Y^{p, q}$ space with wrapped D6 branes
in type IIA, and two, the lift of the resolved $Y^{p, q}$ space with fluxes but without branes again in type IIA. Additionally these two $G_2$
structure manifolds should be related by a flop transition, similar to the one constructed for the $T^{1,1}$ case in \cite{amv}.

\vskip.1in

\noindent If all the three ingredients discussed above are present
 then one would be able to describe geometric transition using the resolved and the
deformed $Y^{p, q}$ manifolds via the duality map given in {\bf figure 1}. In the following we will describe a possible realisation of these
scenarios. Our starting point would be the resolution and the deformation of the cones over
 $Y^{p, q}$ manifolds, which lie in the heart of these scenarios.

\subsection{Resolution and deformation of the cones over $Y^{p,q}$ \label{resolY}}

A natural question is
whether there can be resolutions for the cone over $Y^{p,q}$ as the
resolved conifold. The answer is in the affirmative and
the metric on the resolved cone over $Y^{p,q}$
was obtained explicitly in \cite{Oota:2006pm}, \cite{Lu:2006cw} and
\cite{Martelli:2007pv}. The metric is,
\begin{eqnarray}\label{rsm}
ds^2_{RS}&=&\frac{(1-y)(1-x)}{3}( d\theta^2+\sin^2\theta
d\phi^2)+\frac{(y-x)(1-y)}{h(y)}dy^2+\frac{(x-y)(1-x)}{f(x)}dx^2\nonumber\\
&+&\frac{f(x)}{9(1-x)(x-y)}\Big[d\psi-\cos\theta
d\phi+y(d\beta+\cos\theta d\phi)\Big]^2\nonumber\\
&+&\frac{h(y)}{9(1-x)(y-x)}\Big[d\psi-\cos\theta
d\phi+x(d\beta+\cos\theta d\phi)\Big]^2,
\end{eqnarray}
where $f(y)=2y^3-3y^2+a$ and $h(x)=2x^3-3x^2+b$. We will also take the sechsbein $e^a$ to be the
ones given in eq (2.8) of \cite{pandoman} with appropriate redefinations of the variables therein.

As explained in the subsection \ref{YpqReview}, $y_- < y < y_+$. One can take $x$
to be non-compact and denote two consecutive roots of $h(x)$ by $x_-$
and $x_+$. We focus on the case where the resolution is obtained by
blowing up a $CP^1$, referred to as small partial resolutions
in \cite{Martelli:2007pv}. For this type of resolution we have
$x_-=y_-$ which requires $a=b$. Thus we get,
\begin{eqnarray}
-\infty <x <y_-, \quad y_-< y < y_+, \quad a=b.
\end{eqnarray}
If one takes $x=-r^2/2$ and expand the metric \eqref{rsm} in the large
$r$ it becomes $ds^2_{RS}\rightarrow dr^2+r^2ds^2$ where $ds^2$ is
exactly \eqref{canm}, so it is a cone over $Y^{p,q}$.

Having got the resolution of the cone over $Y^{p, q}$,
we now want to study the deformation of the cone over $Y^{p,q}$, which
should be a mirror of the resolved cone over $Y^{p,q}$. Strominger,
Yau, and Zaslow conjectured that the mirror manifold can be obtained
by three T-dualities \cite{syz}. There are three isometric directions $\psi$,
$\beta$ and $\phi$, so we will first do T-dualities along these
directions. The metric we get after three T-dualities is:
\begin{eqnarray}\label{rst}
ds^2_{\rm SYZ}&=&\frac{(1-y)(1-x)}{3}( d\theta^2+\sin^2\theta
d\phi^2)+\frac{(y-x)(1-y)}{h(y)}dy^2+\frac{(x-y)(1-x)}{f(x)}dx^2\nonumber\\
&+&\frac{f(x)h(y)(x-y)\cos\theta}{9(f(x)y(1-y)^2)-h(y)x(1-x)^2}\Big(d\psi+\frac{f(x)y^2(1-y)-h(y)x^2(1-x)}{f(x)y(1-y)-h(y)x(1-x)}d\beta
+\frac{d\phi}{\cos\theta}\Big)\nonumber\\
&+&\frac{9h(y)(1-x)}{f(x)h(y)(x-y)\cos^2\theta}\Big[(1-x)\cos\theta
d\beta +xd\phi\Big]^2\nonumber\\
&+&\frac{9f(x)(1-y)}{f(x)h(y)(y-x)\cos^2\theta}\Big[(1-y)\cos\theta
d\beta +yd\phi\Big]^2.
\end{eqnarray}
The above metric however cannot be the full answer as T-dualities {\it a la} \cite{syz} require us to take the base to be very large.
In \cite{chen1} (see also \cite{gtpaper}) we saw that making the base large actually {\it mixes} the isometry directions, leading eventually to the
generation of
additional cross-terms missing from the metric obtained by making naive T-dualities. Thus the actual mirror metric will have cross-terms
in addition to what we already have in \eqref{rst}.

The complete picture is rather involved as the recipe for making the base bigger using coordinate transformations {\it a la} \cite{chen1}
is not readily available now. However despite this obstacle,
one thing is clear from the analysis of \cite{chen1}: the resultant metric will not be a K\"ahler manifold, in fact, it may not even be a
complex manifold.
This
is consistent with the result of \cite{altmann1, altmann2} (see also \cite{yaug} where certain obstructions to the existence of Sasaki-Einstein
metrics on this manifold is shown).
It will also be interesting to compare our result with the one got in \cite{Burrington:2005zd}.

\subsection{D5 branes on the resolved $Y^{p, q}$ manifold \label{D5resolY}}

The technical obstacle that we encountered in the previous subsection doesn't prohibit us to write the metric of $N$ D5 branes wrapped
on the two-cycle of the resolved cone over $Y^{p, q}$ manifold. Recently the NS5 brane picture has been studied in \cite{pandoman}. The
analysis of \cite{pandoman} is similar in spirit to the one discussed in \cite{chen1}, both the analyses being motivated by the
work of \cite{MM}. The complete background for $N$ D5 branes wrapped on the resolution two-cycle is given by:
\bg\label{susybg}
&& F_3 = h~{\rm cosh}~\beta~e^{-2\phi} \ast d\left(e^{2\phi} J\right), ~~~~~~~
H_3 = -hF_0^2{\rm sinh}~\beta~e^{-2\phi} d\left(e^{2\phi} J\right)\nonumber\\
&& F_5 = -{1\over 4} (1 + \ast) dA_0 \wedge dx^0 \wedge dx^1 \wedge  dx^2 \wedge  dx^3\\
&& ds^2 = F_0 ds^2_{0123} + \sum_{a = 1}^6 f_a e^{2a}, ~~~~~~ \phi =  {\rm log} ~F_0 + {1\over 2} {\rm log}~ h,\nonumber
\nd
where $e^a$ are the sechsbein defined in \cite{pandoman} and $J$ is the fundamental form associated with the
internal metric. The above background
is supersymmetric by construction and since the RR three-form
$F_3$ is not closed, it represents precisely the IR configuration
of wrapped D5-branes on warped non-K\"ahler resolved $Y^{p, q}$ manifold. The two warp factors ($h, F_0$) as well as the coefficients
$f_a$ in the internal metric are all functions of ($r, y, x$) which, in turn, preserve the three isometries of the internal
space. Notice also that the background has a non-trivial dilaton, with the internal space being a non-K\"ahler resolved cone over $Y^{p, q}$.
The form of the background \eqref{susybg} is similar to the one that we had in \cite{chen1} except now the internal space is different.
This is of course expected if one had to preserve ${\cal N} = 1$ supersymmetry. The five-form, which is switched on to preserve the
susy, has the form $F_5$ in \eqref{susybg} with:
\bg\label{a0def}
A_0 ~&=~ &{{\rm cosh}~\beta~{\rm sinh}~\beta (1-e^{-2\phi}h^{-2}F_0^{-4})
\over e^{2\phi} h^{-2} F_0^{-4}{\rm cosh}^2\beta - {\rm sinh}^2\beta}\nonumber\\
&= ~& (F_0^2 -1 ){\rm tanh}~\beta \left[1 + \left({1-F_0^2\over F_0^2}\right){\rm sech}^2\beta +
\left({1-F_0^2\over F_0^2}\right)^2 {\rm sech}^4 \beta\right].
\nd
Let us now make a few observations.
The parameter $\beta$ that we have in the background is in general constant and could take any value. This means that there is a class of
allowed backgrounds satisfying the supersymmetry condition. Imagine also that we define a six-dimensional internal space in the following way:
\bg\label{jhogra}
ds^2_6 ~ = ~ \left({N F_0 {\rm cosh}^2\beta\over 1+F_0^2 {\rm sinh}^2\beta}\right) \sum_{a = 1}^6 f_a e^{2a},
\nd
then one could easily argue that there are a series of
dualities\footnote{Starting with the background \eqref{simpbg}, we perform a S-duality that transforms the NS three-form to RR three-form
$F_3$ and converts the dilaton $\Phi$ to $\phi$ without changing the metric in the Einstein frame.
We now make three T-dualities along the spacetime directions $x^{1, 2, 3}$ that takes us to type
IIA theory. Observe that this is {\it not} the mirror construction. We then
lift the type IIA configuration to M-theory and perform a boost (with a parameter $\beta$) along the eleventh direction.
This boost is crucial in generating D0-brane {\it gauge} charges in M-theory.
A dimensional reduction back to IIA theory does exactly what we wanted: it generates the necessary
number of D0-brane charges from the boost, without breaking the underlying supersymmetry of the system. Finally,
once we have the IIA configuration, we go back to type IIB by performing the three T-dualities
along $x^{1, 2, 3}$ directions. From the D0-brane charges, we get back our three-brane charges namely the five-form. The
duality cycle also gives us NS three-form $H_3$ as well as the expected RR three-form $F_3$. Therefore the final
configuration is exactly what we required for IR geometric transition: wrapped D5s with necessary sources on a
non-K\"ahler globally defined resolved $Y^{p, q}$ background \eqref{susybg}.
Also as expected, the background preserves
supersymmetry and therefore should be our starting point. One may also note that the thee-forms that we get in \eqref{susybg} satisfy
$$\cosh\beta~H_3 ~+~ F_0^2~\sinh\beta \ast F_3 ~ = ~ 0$$ which is the modified ISD (imaginary self-duality) condition.
For more details, see \cite{MM, chen1}.}
that would convert the following background
\bg\label{simpbg}
ds^2 = ds^2_{0123} + N ds^2_6, ~~~~~ H_{\rm NS} = e^{-2\Phi} \ast d\left(e^{2\Phi} J\right), ~~~~~\Phi = -\phi
\nd
to the one given earlier in \eqref{susybg}. The above background \eqref{simpbg} is of course the one studied in \cite{pandoman}. Although this is
no big surprise, but
it is satisfying to see that our picture can be made consistent with both \cite{pandoman} as well as
\cite{chen1}.

\subsection{Toward geometric transitions for $Y^{p, q}$ manifolds \label{GT}}

Once we have the background \eqref{susybg} and \eqref{a0def} we should be able to use directly the duality cycle
shown in {\bf figure 1}.
This however will turn out to be more subtle than the story that we developed in \cite{chen1}. But before we go about elucidating the issues,
let us clarify certain things about {\it generalized} SYZ. The original work of SYZ \cite{syz} is based on two facts: (a) all Calabi-Yau manifolds
can be written in terms of a $T^3$ fibration over a base ${\cal B}$, and (b) in the limit where ${\cal B}$ is much larger than the $T^3$ fiber,
mirror of the given CY manifold is given by three simultaneous T-dualities along the $T^3$ fiber directions.

For our case, the starting manifold \eqref{susybg} is not a CY manifold but instead
is a six-dimensional manifold with an $SU(3)$ structure and torsion $H_3$.
For this case there does exist a generalisation of the SYZ technique: it is again given by three T-dualities along the 
$T^3$ fiber \cite{micu, minatom, tom}.
The difference now is that we cannot claim that {\it all} $SU(3)$ structure manifolds can be expressed in terms of $T^3$ fibrations
over some base manifolds (although \cite{gross, minatom, tom} has discussed more generic 
cases by applying {\it local} T-dualities). This generalization of the
SYZ technique is called the generalized mirror rule\footnote{For more details as to why the generalized mirror rule would lead to another $SU(3)$
structure manifold that is the {\it mirror} of the original manifold is discussed in \cite{minatom, tom}.}.

Our method now would be to use the generalized SYZ technique to go to the type IIA mirror manifold with wrapped D6-branes. Unfortunately now there are
two subtleties that make the analysis much more non-trivial than the one that we had in \cite{chen1}. The first one is already been discussed
earlier: we don't know exactly what kind of coordinate transformations we should do to make the base bigger than the $T^3$ fiber. Recall that
in \cite{chen1}, out of infinite possible coordinate transformations available,
we could find a class of transformations that can not only make the base bigger but also
lead us to the right mirror manifold. The main reason why we could find that particular class of
transformations earlier was solely based on the fact that we {\it knew} the
existence of a deformed conifold solution. This privileged information, unfortunately, is not available to us
now.

The second issue is even more non-trivial. Looking at the background \eqref{susybg} and from $H_3 = dB_{\rm NS}$, 
we see that the $B_{\rm NS}$ fields will
have components that are {\it parallel} to the directions of the $T^3$ fiber. T-dualities with $B_{NS}$ fields along the directions of duality
lead to non-geometric manifolds! Therefore the type IIA dual manifold will most likely be a non-geometric space which in turn means that the
duality cycle depicted in {\bf figure 1} cannot be very 
straightforward\footnote{There is a third subtlety that has to do with the size of the $T^3$ fiber in the mirror manifold. If the size of the 
fiber is small i.e of ${\cal O}(\alpha')$, 
then supergravity description may not be possible, and one might have to go to a Gepner type sigma model description. For the model 
studied in \cite{gtpaper, chen1} this was not an issue because we could study a class of manifolds parametrised by choice of warp factors 
that not only satisfy 
EOMs but also lie in subspaces, where sugra descriptions are valid, 
on both sides of figure 3 in \cite{chenhet}. These subspaces 
are related by geometric transitions. For generic choices of the warp factors in \cite{chen1, chenhet}, it would be interesting to see if the 
subspaces could incorporate the $Y^{p, q}$ manifolds.}.    

Existence of non-geometric space, however,
does not mean that there is no underlying gauge/gravity duality. In fact in the geometric transition set-up there
were already indications, even for the simplest resolved conifold case, that the full gauge/gravity duality will involve non-geometric manifolds
\cite{halmagi}, although we argued in \cite{chen1} that there is small configuration space of fluxes where we expect the duality to be captured
by purely geometric manifolds. The question now is whether such a scenario, with only geometric spaces, could be realised for the present case also.
We will leave this for future work.

\section{Conclusion and open question \label{conclusion}}

In this paper we studied the scalar spectrum of $Y^{p,q}$ manifold. Earlier works in this direction \cite{Kihara:2005nt, Oota:2005mr} mostly studied
the lowest eigenmodes of the scalar Laplacian, as finding the exact eigenmodes in closed form
for the full tower of states is practically impossible. The main difficulty
lies in the existence of four regular singular points for an operator of Heun type that the scalar Laplacian can be reduced to. Despite this problem
we have managed to find both upper and lower bounds for all the eigenmodes $\lambda_k$ of the scalar Laplacian. Our result can be expressed as:
\bg
\Lambda_k^{[-]} ~ \leq ~ \lambda_k ~ \leq~ \Lambda_k^{[+]}
\nd
where $k = 1, 2,...$ and $\Lambda_k^{[\pm]}$ are given in \eqref{sbound2}. We also show that asymptotically, i.e for large $k$, the eigenmodes grow
quadratically as in \eqref{asymb}. Note that this is the opposite regime of the spectrum of \cite{Kihara:2005nt}, where they give exact lowest eigenvalues.
Our analysis presented here
works best for $a \ll  1$ or equivalently $q \ll p$. By comparing against the known low-lying scalar spectrum for example those 
from \cite{Kihara:2005nt}, we may learn something useful about the bound. Furthermore, our natural guess would be that the 
spectra contain all the BPS and non-BPS states. 
As we saw earlier in \eqref{asymb}, for 
large $k$, the masses are proportional to $k$ and are therefore additive to leading order. 
However for small $k$ we don't have the precise behavior and therefore 
cannot pin-point their BPS or non-BPS nature. In fact, as we wrote above, we expect both these states to show up. More details on this will be 
discussed elsewhere.    
It will be also interesting to study the implications of the consistent massive truncation
along the line of \cite{dallagata, 1003.5374, varela1, varela2} where they also focus on the lowest massive modes. 

The absence of solution in closed form signifies the possibility of introducing numerical techniques to solve the problem. This is
along the lines of \cite{Bachas:2011xa} where a numerical study was done for the simplest Sasaki-Einstein manifold, namely the $S^5$. Unfortunately the
success of the $S^5$ case doesn't necessarily guarantee the same
for the $Y^{p, q}$ case, again precisely due to the fact that the corresponding Heun equation
has four regular singular points. Therefore it seems at this stage, unless we know how to tackle these singularities, a closed form solution via
numerical analysis looks unfeasible. Any progress in this direction will be a productive boost for completing the duality
dictionary for this 
case\footnote{In ~\cite{Kihara:2005nt} the authors used AdS/CFT correspondence to
map the Reeb Killing vector $Q_R$, $\hat{K}=J(J+1)$ and $pN_{\alpha}$
to the R-symmetry, $SU(2)$ spin and $U(1)$ flavor charge 
on the field theory side respectively. They compared states with
quantum numbers ($Q_R$, $J$, $N_{\alpha}$) and chiral operators with
the same charges under these symmetries. With some chosen values of $Q_R$,
$J$, $pN_{\alpha}$ they found the eigenvalues for the scalar
Laplacian of $Y^{p,q}$ and from there claimed that the ground states satisfy
the BPS condition. We find that the eigenvalues $N_{\psi}$ and $pN_{\alpha}$ are not always integers, and their results are covered
in our spectrum with some specified values of $m$, $n$, $l$, $j$. This 
can also be used to determine the range of the constant $C$ \label{fang1}.}. 
For example computing the super-conformal index along the lines of \cite{rastelli}, or even going beyond the 
supergravity modes {\it a-la} \cite{pandozayas}.  

Another line of thought that we followed in this paper is the non-conformal extensions of the conformal examples. These non-conformal models are
closer to the geometric transition models of \cite{vafaGT, gtpaper, chen1} and therefore would require the existence of the corresponding deformed
cones over the $Y^{p, q}$ manifolds. The deformed cones over $Y^{p, q}$ manifolds, unfortunately,
couldn't be Calabi-Yau manifolds \cite{altmann1, altmann2, yaug} so the
underlying picture cannot be as simple as the ones studied in \cite{gtpaper, chen1}. Our analysis, however,
reveals that the gravity duals might not even be
geometric manifolds, so that the obstructions pointed out in \cite{altmann1, altmann2, yaug} could be circumvented. Although no concrete
examples exist at this stage, the above approach is a hopeful avenue to realise non-conformal duals. Additionally a success in this direction
would also be a good test for the generalized mirror symmetry that relate two manifolds with $SU(3)$ structures (i.e manifolds with intrinsic
torsions and $H_3$ fluxes).

Clearly what we opened up here is just the tip of an iceberg, and happily there are more questions than answers right now. In future works
we will address some of these issues in more details.

\vskip.2in

\centerline{\bf Acknowledgements}

\vskip.1in

\noindent It is a great pleasure to thank James Sparks, Phillip Szepietowski, David Simmon-Duffin, and Shing-Tung Yau for
helpful correspondence and discussions. K. D would like to thank Ruben Minasian and Alessandro Tomasiello for helpful correspondences.
J. S would like to thank Amihay Hanany, Sungjay Lee,
Daniel Robbins, Andrew Royston, and Jock McOrist for discussions. We would also like to thank the referee for his comments that helped us to 
improve the contents of our paper.
The work of F.C, K. D and J. S is supported in part by NSERC grants, the work of N. K is supported in part by NSERC grant RGPIN 105490-2011
and the work of A. E is supported in part by the MICINN and the UCM-Banco
Santander under grants number FIS2008-00209 and GR58/08-910556.


\appendix

\section{Eigenvalues of the differential operator $S$ \label{eigenAppen}}
To make this paper self-contained, here we will borrow two lemmas proved in \cite{EK10}.

Let us consider the differential operator
\begin{equation}\label{S}
  S_{ml}(\La):=\frac1{\rho(y)}\frac\pd{\pd y}\rho(y)\,w(y)\,r(y)\,\frac\pd{\pd y} -\frac1{w(y)}\bigg(\frac{\si l}\tau\bigg)^2- \frac9{r(y)}\bigg(2m-h(y)\,\frac{\si l}\tau\bigg)^2
-\frac{6\La}{1-y}\,,
\end{equation}
arising from~\eqref{jbaaz}, which depends on
a real parameter $\La\ge 0$. It is clear that we cannot hope to
express the solutions of the formal eigenvalue equation
\begin{equation}\label{eqS}
  S_{ml}(\La)\,w=-\la w
\end{equation}
in closed form using special functions because~\eqref{eqS} is a
Fuchsian differential equation with {\it four} regular singular
points, located at the three roots of the cubic~ $a - 3y^2 + 2y^3 = 0$, at
$1$, at $\pm a^{1/2}$ and at infinity\footnote{An ordinary differential equation whose only singular points, including the point at infinity, are regular singular points is called a Fuchsian ordinary differential equation.}. However, the information
contained in the following lemma will suffice for our purposes.

\noindent {\bf Lemma 1}:
For all $\La\ge 0$, the differential operator~\eqref{S} defines a
nonnegative self-adjoint operator in $L^2((y_-,y_+),\rho(y)\,dy)$,
which we also denote by $S_{ml}(\La)$, whose domain consists of the
functions $w\in {\cal AC}^1((y_-,y_+))$ such that $S_{ml}(\La)w\in
L^2((y_-,y_+))$ and
 \begin{equation*}
\lim_{y\searrow y_-}y\,w'(y)=0\; \text{ if }m=(2p-q)\si l/4 \quad\text{ and }\quad \lim_{y\nearrow y_+}y\,w'(y)=0\; \text{ if }m=-q\si l/4\,.
\end{equation*}
Its spectrum consists of a decreasing sequence of eigenvalues
$(-\ell_{mlk}(\La))_{k\in\NN}\searrow-\infty$ of multiplicity one
whose associated normalized eigenfunctions $w_{mlk}(\La)$ are
${\cal O}((y_+-y)^{|m+q\si l/4|})$ as $y\nearrow y_+$ and
${\cal O}((y-y_-)^{|m+(q-2p)\si l/4|})$ as $y\searrow y_-$.

\noindent {\bf Proof}:
Let $y_\epsilon $ be one of the endpoints of the interval $(y_-,y_+)$ and
set $\zeta:=y-y_\epsilon$. An easy computation shows that
  \[
a-3y^2+2y^3=-6y_\ep(1-y_\ep)\,\zeta+\cO(\zeta^2)\,,\qquad r(y)=-\frac\zeta{3y_\ep}+\cO(\zeta^2)
  \]
  as $y\to y_\ep$, which shows that the differential equation~\eqref{eqS} can be asymptotically written as
  \[
-\big(12y_\ep\,\zeta+\cO(\zeta^2)\big)\,\tw''(\zeta)-\big(12y_\ep+\cO(\zeta)\big)\, \tw'(\zeta)+\Bigg[ \frac{3y_\ep}\zeta\bigg(2m-h(y_\ep)\frac{\si l}\tau\bigg)^2+\cO(1)\Bigg]\,\tw(\zeta)=0\,,
  \]
with $\tw(\zeta):=w(\zeta+y_\ep)$ standing for the expression of the function $w(y)$ in the new variable~$\zeta$.

It then follows that the characteristic exponents of the
equation~\eqref{eqS} at $y_\ep$ are $\pm\nu_\ep$, with
$\nu_\ep:=|m-h(y_\ep)\si l/(2\tau)|$. Using
\begin{equation}\label{pq}
  \frac{h(y_+)-h(y_-)}{2\,h(y_+)}=\frac{p}{q}\,, \qquad
  \tau \equiv -2\,h(y_+)/q\,,\qquad \si:=\lcm\{2,pq,2p-q\}\,.
\end{equation}
one can immediately derive the more manageable
formula
  \begin{equation}\label{nuep}
\nu_+=\big| m+q\si l/4\big|\,,\qquad \nu_-=\big|m+(q-2p)\si l/4\big|\,.
\end{equation}
Let us now consider the following lemma.

\noindent{\bf Lemma 2}:
Let $\cY$ be the complex Hilbert space
\[
\cY:=\Big\{\big(u_{nml}\big)_{n,m,l\in\ZZ}:
u_{nml}\in L^2\big((y_-,y_+),\rho(y)\,dy\big)\otimes L^2\big((0,\pi),\sin\theta\,d\theta\big)\Big\}\,,
\]
endowed with the norm
\[
 \Big\|\big(\Phi_{nml}\otimes\Theta_{nml}\big)_{n,m,l\in\ZZ}\Big\|_{\cY}^2   :=  \sum_{n,m,l\in
  \ZZ}\bigg(\int_{y_-}^{y_+} \big|\Phi_{nml}(y)\big|^2 \rho(y)
\,dy\bigg)\bigg(\int_0^{2\pi}\big|\Theta_{nml}(\theta)\big|^2 \sin\theta \,d\theta\bigg)\,,
\]
and with $\tau$ and $\si$ defined as in \eqref{pq}.
Then the map defined by
\begin{equation}\label{Y}
\cY\ni\big(\Phi_{nml}\otimes\Theta_{nml}\big)_{n,m,l\in\ZZ}\mapsto
\sum_{n,m,l\in\ZZ}
\Phi_{nml}(y)\,\Theta_{nml}(\theta)\,\frac{\e^{\I(n\phi+2m\psi+\si
l\al/\tau)}}{(2\pi)^{3/2}}\in L^2(Y^{p,q})\,,
\end{equation}
defines an isomorphism between $\cY$ and $L^2(Y^{p,q})$.

Since $\si$ is even by the above lemma, it stems from the latter
equation that $2\,\nu_\ep$ is a nonnegative integer. Therefore, it
is standard that the symmetric operator defined by~\eqref{S} on
$C^\infty_0((y_-,y_+))$ is in the limit point case at $y_\ep$ if and
only if $\nu_\ep\neq0$. If $\nu_+\nu_-\neq0$, the latter operator is
then essentially self-adjoint on $C_0^\infty((y_-,y_+))$, and has a
unique self-adjoint extension of domain~\cite{DS88}
\begin{equation*}
\cD:=\big\{w\in {\cal AC}^1((y_-,y_+)):S_{ml}(\La)\,w\in L^2((y_-,y_+))\big\}\,..
\end{equation*}
When $\nu_+\nu_-=0$, the above symmetric operator is not essentially
self-adjoint. In this case, in order to rule out logarithmic
singularities we shall choose its Friedrichs extension~\cite{MZ00},
whose domain consists of the functions $w\in\cD$ such that
\[
\lim_{y\searrow y_-}y\,w'(y)=0\;\text{ if }\nu_-=0\quad\text{ and }\quad \lim_{y\nearrow y_+}y\,w'(y)=0\;\text{ if }\nu_+=0\,,
\]
It is well known~\cite{DS88} that $S_{ml}(\La)$ is then a
nonnegative operator with compact resolvent and that its eigenvalues
are nondegenerate.

\newpage

\bibliographystyle{JHEP}
\bibliography{Ypq}

\end{document}